\documentclass[twocolumn]{aastex631}

\newcommand{\um}{\,$\,\mu$m}
\newcommand{\sqdeg}{\,\,deg$^2$}
\newcommand{\m} {\,\,m}
\newcommand{\cm}{\,\,cm}
\newcommand{\days}{\,\,days}
\newcommand{\hr}{\,\,hr}
\newcommand{\km}{\,\,km}
\newcommand{\degC}{$^\circ$\,C}
\begin{document}

\title{Cryoscope: A Cryogenic Infrared Survey Telescope in Antarctica}

\author[0000-0002-5619-4938]{Mansi M. Kasliwal}
\affil{Cahill Center for Astrophysics, California Institute of Technology, Pasadena, CA 91125, USA}

\author[0000-0001-6627-9903]{Nicholas Earley}
\affil{Cahill Center for Astrophysics, California Institute of Technology, Pasadena, CA 91125, USA}

\author[0000-0001-7062-9726]{Roger Smith}
\affil{Cahill Center for Astrophysics, California Institute of Technology, Pasadena, CA 91125, USA}

\author[0000-0002-7188-8428]{Tristan Guillot}
\affil{Universit\'e C\^ote d’Azur, Observatoire de la C\^ote d’Azur, CNRS, Laboratoire Lagrange, CS 34229, F-06304 Nice Cedex 4, France}

\author[0000-0001-9304-6718]{Tony Travouillon}
\affil{Research School of Astronomy and Astrophysics, Australian National University, Canberra, ACT 2611, Australia}

\author{Jason Fucik}
\affil{Cahill Center for Astrophysics, California Institute of Technology, Pasadena, CA 91125, USA}

\author[0000-0002-0856-4527]{Lyu Abe}
\affil{Universit\'e C\^ote d’Azur, Observatoire de la C\^ote d’Azur, CNRS, Laboratoire Lagrange, CS 34229, F-06304 Nice Cedex 4, France}

\author{Timothee Greffe}
\affil{Cahill Center for Astrophysics, California Institute of Technology, Pasadena, CA 91125, USA}

\author[0000-0001-7948-6493]{Abdelkrim Agabi}
\affil{Universit\'e C\^ote d’Azur, Observatoire de la C\^ote d’Azur, CNRS, Laboratoire Lagrange, CS 34229, F-06304 Nice Cedex 4, France}

\author[0000-0003-1412-2028]{Michael C. B. Ashley}
\affil{School of Physics, University of New South Wales, Sydney NSW 2052, Australia}

\author[0000-0002-5510-8751]{Amaury H.M.J. Triaud}
\affil{School of Physics and Astronomy, University of Birmingham, Edgbaston, Birmingham B15 2TT, United Kingdom}

\author[0000-0002-1481-4676]{Samaporn Tinyanont}
\affil{National Astronomical Research Institute of Thailand, 260 Moo 4, Donkaew, Maerim, Chiang Mai 50180, Thailand}

\author[0000-0002-7686-3334]{Sarah Antier}
\affil{Universit\'e C\^ote d’Azur, Observatoire de la C\^ote d’Azur, Laboratoire Artemis, Nice, France}
\affiliation{IJCLab, Univ Paris-Saclay, CNRS/IN2P3, Orsay, France}

\author[0000-0002-4278-1437]{Philippe Bendjoya}
\affil{Universit\'e C\^ote d’Azur, Observatoire de la C\^ote d’Azur, CNRS, Laboratoire Lagrange, CS 34229, F-06304 Nice Cedex 4, France}

\author[0009-0007-0323-4733]{Rohan Bhattarai}
\affil{Cahill Center for Astrophysics, California Institute of Technology, Pasadena, CA 91125, USA}

\author{Rob Bertz}
\affil{Cahill Center for Astrophysics, California Institute of Technology, Pasadena, CA 91125, USA}

\author{James Brugger}
\affil{Cahill Center for Astrophysics, California Institute of Technology, Pasadena, CA 91125, USA}

\author[0000-0001-9892-2406]{Artem Burdanov}
\affil{Department of Earth, Atmospheric and Planetary Science, Massachusetts Institute of Technology, Cambridge, MA, USA}

\author[0000-0002-4770-5388]{Ilaria Caiazzo}
\affil{Cahill Center for Astrophysics, California Institute of Technology, Pasadena, CA 91125, USA}

\author[0000-0001-5242-3089]{Benoit Carry}
\affil{Universit\'e C\^ote d’Azur, Observatoire de la C\^ote d’Azur, CNRS, Laboratoire Lagrange, CS 34229, F-06304 Nice Cedex 4, France}

\author[0000-0003-2688-7511]{Luca Casagrande}
\affil{Research School of Astronomy and Astrophysics, Australian National University, Canberra, ACT 2611, Australia}

\author[0000-0003-1673-970X]{Brad Cenko}
\affil{Astrophysics Science Division, NASA Goddard Space Flight Center, Mail Code 661, Greenbelt, MD 20771, USA}

\author[0000-0001-5703-2108]{Jeff Cooke}
\affil{Centre for Astrophysics and Supercomputing, Swinburne University of Technology, MN 74, PO Box 218, Hawthorn, VIC 3122, Australia}
\affil{Australian Research Council Centre of Excellence for Gravitational Wave discovery (OzGrav), Australia\\}

\author[0000-0002-8989-0542]{Kishalay De}
\altaffiliation{NASA Einstein Fellow}
\affil{Department of Astronomy and Columbia Astrophysics Laboratory, Columbia University, 550 W 120th St. MC 5246, New York, NY 10027, USA}
\affil{Center for Computational Astrophysics, Flatiron Institute, 162 5th Ave., New York, NY 10010, USA}

\author{Richard Dekany}
\affil{Cahill Center for Astrophysics, California Institute of Technology, Pasadena, CA 91125, USA}

\author[0009-0007-5876-546X]{Vincent Deloupy}
\affil{Universit\'e C\^ote d’Azur, Observatoire de la C\^ote d’Azur, CNRS, Laboratoire Lagrange, CS 34229, F-06304 Nice Cedex 4, France}
\affiliation{\'{E}cole Normale Sup\'erieure, D\'epartement de Physique, Rue d’Ulm, 75005 Paris Cedex 5, France}

\author[0000-0001-5729-1468]{Damien Dornic}
\affil{Aix Marseille University, CNRS/IN2P3, CPPM, Marseille, France}

\author{Lauren Fahey}
\affil{Cahill Center for Astrophysics, California Institute of Technology, Pasadena, CA 91125, USA}

\author{Don Figer}
\affil{Center for Detectors, Rochester Institute
of Technology, Rochester, NY, USA}

\author[0000-0001-6280-1207]{Kenneth Freeman}
\affil{Research School of Astronomy and Astrophysics, Australian National University, Canberra, ACT 2611, Australia}

\author[0000-0002-7197-9004]{Danielle Frostig}
\affil{Center for Astrophysics | Harvard \& Smithsonian, 60 Garden Street, Cambridge, MA 02138, USA}

\author[0000-0002-3168-0139]{Matthew J. Graham}
\affil{Cahill Center for Astrophysics, California Institute of Technology, Pasadena, CA 91125, USA}

\author[0000-0002-3164-9086]{Maximilian Günther}
\affil{European Space Agency (ESA), European Space Research and
Technology Center (ESTEC), Keplerlaan 1, 2201 AZ Noordwĳk,
The Netherlands}

\author{David Hale}
\affil{Cahill Center for Astrophysics, California Institute of Technology, Pasadena, CA 91125, USA}

\author[0000-0001-7516-4016]{Joss Bland-Hawthorn}
\affil{Sydney Institute for Astronomy, School of Physics, A28, The University of Sydney, NSW 2006, Australia}
\affil{ARC Center of Excellence for All Sky Astrophysics in 3 Dimensions (ASTRO 3D), Canberra, ACT 2611, Australia}

\author{Giulia Illuminati}
\affil{Universit\`a di Bologna, Dipartimento di Fisica e Astronomia, v.le C. Berti-Pichat, 6/2,
Bologna, 40127 Italy}

\author[0000-0001-5754-4007]{Jacob Jencson}
\affil{IPAC, Mail Code 100-22, Caltech, 1200 E. California Blvd., Pasadena, CA 91125, USA}

\author[0000-0003-2758-159X]{Viraj Karambelkar}
\affil{Cahill Center for Astrophysics, California Institute of Technology, Pasadena, CA 91125, USA}

\author{Renee Key}
\affil{Centre for Astrophysics and Supercomputing, Swinburne University of Technology, Melbourne, VIC 3122, Australia}

\author[0000-0003-0778-0321]{Ryan M. Lau}
\affil{NSF’s NOIRLab, 950 N. Cherry Avenue, Tucson, 85719, AZ,
USA}

\author[0009-0001-6911-9144]{Maggie Li}
\affil{Cahill Center for Astrophysics, California Institute of Technology, Pasadena, CA 91125, USA}

\author{Philip Lubin}
\affil{Department of Physics, University of California -- Santa Barbara, Santa Barbara, CA 93106, USA}

\author{Don Neill}
\affil{Cahill Center for Astrophysics, California Institute of Technology, Pasadena, CA 91125, USA}

\author{Rishi Pahuja}
\affil{Cahill Center for Astrophysics, California Institute of Technology, Pasadena, CA 91125, USA}

\author[0000-0001-8646-4858]{Elena Pian}
\affil{INAF,  Astrophysics and Space Science Observatory, Via Gobetti 101, Bologna, I-40129, Italy}

\author[0000-0001-7717-5085]{Antonio de Ugarte Postigo}
\affil{Aix Marseille Univ, CNRS, CNES, LAM Marseille, France}

\author{Mitsuko Roberts}
\affil{Cahill Center for Astrophysics, California Institute of Technology, Pasadena, CA 91125, USA}

\author{Hector Rodriguez}
\affil{Cahill Center for Astrophysics, California Institute of Technology, Pasadena, CA 91125, USA}

\author[0000-0003-4725-4481]{Sam Rose}
\affil{Cahill Center for Astrophysics, California Institute of Technology, Pasadena, CA 91125, USA}

\author[0000-0002-4794-6835]{Ashley J. Ruiter}
\affil{School of Science, University of New South Wales Canberra, Australia Defence Force Academy, ACT 2600, Australia}
\affil{ARC center of Excellence for All Sky Astrophysics in 3 Dimensions (ASTRO 3D), Canberra, ACT 2611, Australia}

\author[0000-0003-3914-3546]{François-Xavier Schmider}
\affil{Universit\'e C\^ote d’Azur, Observatoire de la C\^ote d’Azur, CNRS, Laboratoire Lagrange, CS 34229, F-06304 Nice Cedex 4, France}

\author[0000-0003-3769-9559]{Robert A. Simcoe}
\affil{Department of Physics and Kavli Institute for Astrophysics and Space Research, Massachusetts Institute of Technology, 77 Massachusetts Ave, Cambridge, MA 02139, USA}

\author[0000-0003-2434-0387]{Robert Stein}
\affil{Department of Astronomy, University of Maryland, College Park, MD 20742, USA}
\affil{Joint Space-Science Institute, University of Maryland, College Park, MD 20742, USA} 
\affil{Astrophysics Science Division, NASA Goddard Space Flight Center, Mail Code 661, Greenbelt, MD 20771, USA} 

\author[0000-0002-3503-3617]{Olga Suarez}
\affil{Universit\'e C\^ote d’Azur, Observatoire de la C\^ote d’Azur, CNRS, Laboratoire Lagrange, CS 34229, F-06304 Nice Cedex 4, France}

\author[0000-0002-3958-0343]{Edward N. Taylor}
\affil{Centre for Astrophysics and Supercomputing, Swinburne University of Technology, Melbourne, VIC 3122, Australia}

\author{Bob Weber}
\affil{Cahill Center for Astrophysics, California Institute of Technology, Pasadena, CA 91125, USA}

\author[0000-0001-7987-295X]{Linqing Wen}
\affil{Department of Physics, University of Western Australia, Perth, Western Australia 6009, Australia}

\author[0000-0003-2415-2191]{Julien de Wit}
\affil{Department of Earth, Atmospheric and Planetary Science, Massachusetts Institute of Technology, Cambridge, MA, USA}

\author{Ray Zarzaca}
\affil{Cahill Center for Astrophysics, California Institute of Technology, Pasadena, CA 91125, USA}

\author{Jake Zimmer}
\affil{Cahill Center for Astrophysics, California Institute of Technology, Pasadena, CA 91125, USA}

\begin{abstract}
We present Cryoscope---a new 50{\sqdeg} field-of-view, 1.2{\m} aperture, $K_{dark}$ survey telescope to be located at Dome C, Antarctica. Cryoscope has an innovative optical-thermal design wherein the entire telescope is cryogenically cooled. Cryoscope also explores new detector technology to cost-effectively tile the full focal plane. Leveraging the dark Antarctic sky and minimizing telescope thermal emission, Cryoscope achieves unprecedented deep, wide, fast and red observations, matching and exceeding volumetric survey speeds from the {\it Ultraviolet Explorer}, Vera Rubin Observatory, {\it Nancy Grace Roman Space Telescope}, {\it SPHEREx}, and {\it NEO Surveyor}. By providing coverage beyond wavelengths of 2{\um}, we aim to create the most comprehensive dynamic movie of the most obscured reaches of the Universe. Cryoscope will be a dedicated discovery engine for electromagnetic emission from coalescing compact binaries, Earth-like exoplanets orbiting cold stars, and multiple facets of time-domain, stellar and solar system science. In this paper, we describe the scientific drivers and technical innovations for this new discovery engine operating in the $K_{dark}$ passband, why we choose to deploy it in Antarctica, and the status of a fifth-scale prototype designed as a Pathfinder to retire technological risks prior to full-scale implementation. We plan to deploy the Cryoscope Pathfinder to Dome C in December 2026 and the full-scale telescope by 2030.
\end{abstract}


\section{Introduction}

The astrophysics community has identified opening new windows into the dynamic universe, spanning all energy regimes of electromagnetic, elementary particle, and gravitational wave detection as high priority \citep[e.g. Astro2020 Decadal Survey,][]{astro2020dec}.  
At the midpoint of the decade, new survey telescopes, driven by novel technologies, are coming online to provide high-cadence, multi-band coverage of the sky. In the optical, the Zwicky Transient Facility \citep{Bellm2019}, ATLAS \citep{tonry:2011PASP..123...58T} and Pan-STARRS1 \citep{Chambers:2016arXiv161205560C} have been operating for several years, and first light with the Vera Rubin Observatory \citep{VRO} is planned for later this year. In the ultraviolet, the {\it Ultraviolet Explorer} \citep{Kulkarni2021} and the {\it ULTRASAT} \citep{ultrasat2024} satellites are slated for launch in this decade. 
In the X-ray and $\gamma$-ray wavelengths, recent launches of the {\it Einstein Probe} \citep{EP2022}, {\it SVOM} \citep{SVOM2022} and {\it SRG} \citep{SRG2021} satellites have opened up the time-domain. In the radio, the
upcoming DSA-2000 \citep{DSA2000} and the Square Kilometer Array \citep{SKA2009} will have survey speeds that are orders of magnitude faster than predecessor surveys. 

Here, we focus on the infrared window beyond 1{\um}, which is key to unveiling the physics of cool star-forming regions, dust-creating transients, or temperate exoplanets. Infrared surveys have been limited in their field-of-view due to detector cost and thermal sky background.

Pioneering infrared surveys include {\it IRAS} \citep{IRAS}, {\it ISO/ISOCAM} \citep{isocam1996}, {\it WISE/NEOWISE} \citep{WISE2010,NEOWISE2011}, 2MASS \citep{2MASS} and VISTA \citep{VISTA}. Launched in the 1980s, {\it IRAS} was the first infrared survey satellite to undertake an all-sky survey, with {\it ISO/ISOCAM} performing detailed observations of selected regions in the 2.5 to 18{\um} bandpass soon after. {\it WISE/NEOWISE}, launched in 2009, provided orders-of-magnitude improvements upon {\it IRAS'} sensitivity and spatial resolution, producing the longest time-baseline catalog in the mid-infrared with repeat observations over more than a decade. In the near-infrared $J$ (1.25{\um}), $H$ (1.65{\um}), and $K_s$ (2.16{\um}) bandpasses that are observable from the ground, 2MASS and VISTA produced single reference surveys rather than high-cadence data. 2MASS, using two 1.3{\m} telescopes, operated between 1997 and 2001, generating an all-sky release of 471 million sources with high photometric accuracy. VISTA/VIRCAM, using a 4{\m} telescope, extended this catalog with deeper and higher-resolution observations, focusing on specific areas of the sky, and is now decommissioned.

New facilities are now emerging to produce high-cadence imaging surveys of the infrared sky. The first, commissioned in 2018, is the Palomar-Gattini-IR telescope \citep{PGIR}. The 30{\cm} telescope has a 25{\sqdeg} field-of-view and has been surveying the accessible Northern sky (20,000{\sqdeg}) to $15.7~M_{AB}$ ($J$-band) every two nights. The telescope uses a cooled Teledyne H2RG camera with a cutoff at 1.7{\um} to avoid the thermal background at 2{\um}. The operations of the next generation infrared surveyor at Palomar Observatory are also now in full swing: the 1{\m} WINTER telescope (Wide-field Infrared Transient Explorer, \citet{WINTER2024, WINTER}) has a 1.2{\sqdeg} InGaAs camera and reaches a depth of 18.5\,\,mag in the $yJH$-bands. In addition to the Palomar surveyors, the PRIME survey, with a 3{\sqdeg} H4RG camera, has come online at the South African Astronomical Observatory \citep{Yama2023,PRIME2024}. While the WINTER and PRIME telescopes are in operation, a complementary surveyor is being built in Australia. This observatory, the Dynamic REd All-sky Monitoring Survey (DREAMS) is due to be commissioned later this year and will provide similarly high-cadenced data of the southern sky \citep{DREAMS}. The 0.5{\m} telescope will feed 6 InGaAs cameras yielding a field-of-view of 4{\sqdeg}. Together, these four telescopes offer full sky coverage of both hemispheres with a combination of depth and cadence that cover many science cases including stellar science, gravitational wave follow-up and extrasolar planetary science. 

These telescopes, however, are limited in sensitivity due to their small apertures and high sky backgrounds. Space directly eliminates sky background. {\it Euclid} \citep{euclid2024}, launched in 2023, and the {\it Nancy Grace Roman Space Telescope}\footnote{\url{https://roman.gsfc.nasa.gov}}, set to launch in 2026, will reach unmatched resolutions and depths in the infrared. However, their fields-of-view ({\emph Roman}'s 0.281{\sqdeg} to {\emph Euclid}'s 0.55{\sqdeg}) and science missions do not make them suitable for dedicated high-cadence, large-area time-domain astronomy.

Wider field-of-view space-based, infrared surveyors will include {\it SPHEREx}\footnote{\url{https://spherex.caltech.edu}} \citep{SPHEREX} and \emph{NEO Surveyor} \citep{mainzer2023} by the end of the decade. {\it SPHEREx} is the latest mid-scale explorer from NASA, which will deliver spectro-photometric coverage from 0.75 to 5{\um} via linear-variable filters over a 40{\sqdeg} total field-of-view ($3.5^\circ \times 3.5^\circ$ for each of its six detector bands). {\it SPHEREx} will not, however, undertake high-cadence exposures of the entire sky or targeted pointings upon alerts, instead delivering all-sky maps four times during its two-year nominal mission. Extending from 4-10{\um}, \emph{NEO Surveyor} with its $\sim 12${\sqdeg} field-of-view, set to launch by the end of the decade,  will build on the legacy of {\it WISE/NEOWISE}, increasing mid-infrared field-of-view and survey speed by an order of magnitude. \emph{NEO Surveyor} will repeatedly scan a region of the ecliptic plane, with ecliptic latitudes limited to $\pm40$ degrees, for its science goals.

A sensitive, high-cadence and wide-field survey is essential to uncovering the dynamic infrared universe. To bridge the current capability gap, we present Cryoscope, an infrared surveyor to be operated robotically at the Concordia Station at Dome C, Antarctica. With a design capable of taking advantage of the low thermal background and seeing conditions of the site, the 1.2{\m} telescope will be able to perform the deepest and fastest survey ever attempted in the infrared. Cryoscope will be able to survey the entire accessible southern sky $\approx$\,20,000{\sqdeg} to a point-source depth of 21.9\,\,mag AB in the $K$-band every 13 hours continuously in the winter. 

\begin{figure}
  \includegraphics{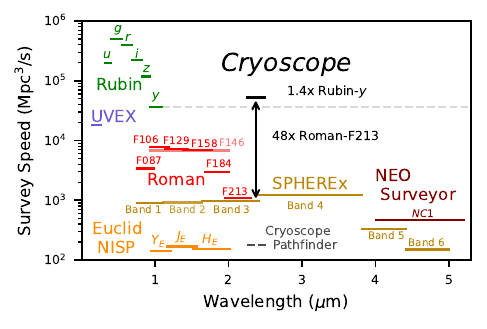}
  \caption{Volumetric survey speed of Cryoscope and the prototype Cryoscope Pathfinder compared to the Rubin Observatory, {\it Ultraviolet Explorer (UVEX)}, the {\emph Roman} space telescope (Galactic bulge and high-latitude time domain surveys), {\it Euclid} space telescope, {\it SPHEREx} all-sky survey, and \emph{NEO Surveyor} up to 5.2{\um}. Refer to Appendix \ref{appendix} for survey parameters in this calculation.
  \label{fig:vspeed}
  }
\end{figure}

\begin{figure}
    \includegraphics[width=\columnwidth]{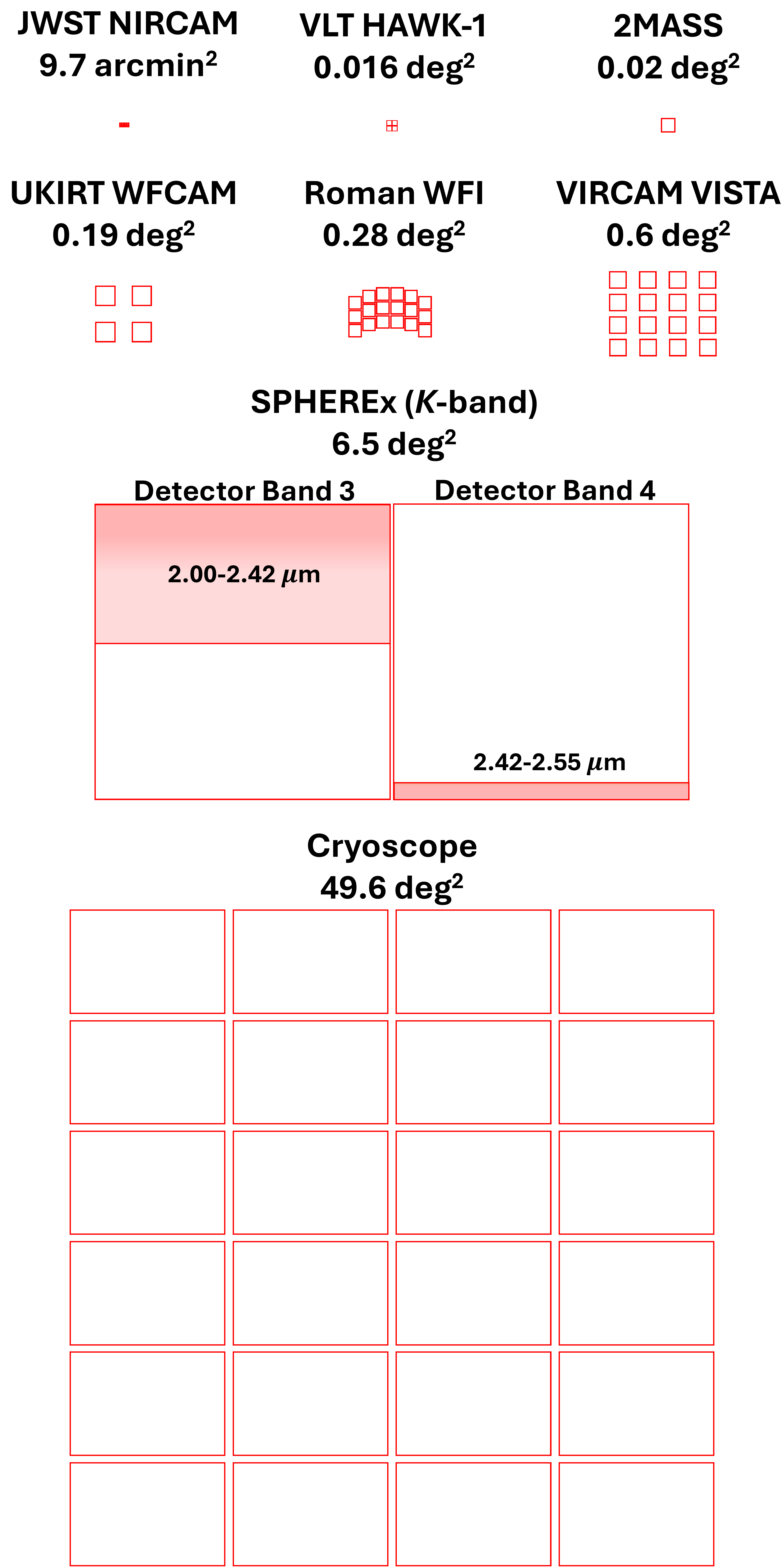}
  
  \caption{Cryoscope's active field-of-view exceeds other surveys in near-infrared wavelengths between 2--2.55{\um}. Given the linear-variable filters (LVF) on {\it SPHEREx}, two of its six detector bands provide coverage in the $K$-band \citep{Hui:2024SPIE13092E..3NH}. We approximate field-of-view in these spectral channels, excluding the slight curvature in the wavelength response across detector bands.
  \label{fig:fov}
  }
\end{figure}

Cryoscope is designed to undertake a time-domain survey that is unparalleled in volumetric mapping speed in the infrared bands (see Figure~\ref{fig:vspeed}). Cryoscope's volumetric mapping speed exceeds the speed of the reddest optical filter, $y$-band, of the Rubin Observatory by a factor of 1.4. Cryoscope's speed is $48\times$ faster than reference surveys with the \emph{Roman Observatory} (e.g., considering the reddest wavelength coverage in the Galactic bulge time domain survey\footnote{\url{https://asd.gsfc.nasa.gov/roman/comm_forum}}).
For the full-scale Cryoscope, the 1.2{\m} aperture combined with the dark Concordia sky, yields a $5\sigma$ point-source depth of 24.1\,\,mag AB achieved in 1{\hr}
(its fifth-scale prototype, the Pathfinder, yields a depth of 20.2\,\,mag in the same time). Cryoscope's 49.6{\sqdeg} field-of-view (see Figure~\ref{fig:fov}) is $5\times$\ larger than Rubin's and $200\times$ larger than {\it Roman}'s. Furthermore, combining this exquisite infrared sensitivity with the continuous observing season in Antarctica enables a unique search of the southern infrared sky, facilitating a vast array of science cases from characterizing temperate and terrestrial planets to discovering explosive transient events and monitoring their light curves.

In this paper, we present the technical innovation in \S\ref{sec:technology}, status of the fifth-scale prototype to retire technological risk in \S\ref{sec:pathfinder}, science drivers in \S\ref{sec:science} and conclude with the path forward in \S\ref{sec:summary}.


\section{Technical Description}\label{sec:technology}
\subsection{The site}

\begin{figure}
	\includegraphics[width=\columnwidth]{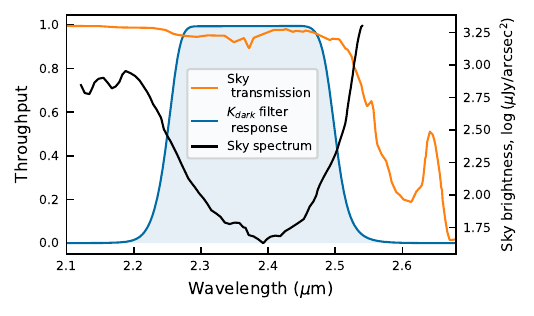}
    \caption{Measured infrared sky background from the South Pole taken from \citet{Ashley1996} and sky transmission from \citet{Hidas2000}. The $K_{dark}$ filter response was optimized to exploit the window between 2.2 and 2.5{\um} over the Antarctic plateau, similar to the methods in \citet{li2016}. The sky brightness at Dome C is expected to be even lower due to its higher elevation and colder temperatures.}
    \label{fig:sky}
\end{figure}

The high Antarctic plateau is indispensable to observational astronomy. With a contiguous area spanning the size of Australia at elevations exceeding 2,000{\m}, the region is home to numerous international research stations. The South Pole station (2,835{\m}) currently hosts an American research program that specializes in sub-millimeter observations and high-energy particle detections. Higher locations such as Dome A (4,093{\m}) and Dome C (3,233{\m}), respectively overseen by China and France/Italy, host observatories operating at optical wavelengths. The Dome C station is particularly attractive for a survey telescope because it has people on-site year-round and is already home to robust observational programs including The Antarctic Search for Transiting Planets (ASTEP) \citep{Guillot+2015,Crouzet2020,Schmider2022,Dransfield2022}. The scientific advantages of operating such facilities at this location are as follows:

\begin{itemize}
    \item  {\bf Low sky background:} As shown in \citet{Ashley1996} and \citet{Nguyen1996}, the $K$-band sky background at the South Pole is about 40 times lower than the best temperate ground sites thanks to a combination of low temperature and low water vapor content. With the minimum at 2.4{\um} caused by the disappearance of airglow lines, the $K_{dark}$ window (2.2–2.5{\um}) is the one to be optimally exploited for astronomy \citep[Figure~\ref{fig:sky}; also see][]{li2016}. At the higher elevations and colder temperatures afforded by Dome C, the sky brightness is anticipated to be even lower than at the South Pole. 

    \item {\bf Low seeing:} Seeing quantifies the degradation of the resolution of astronomical images caused by turbulence when air of different temperature (thus density) mixes together. The seeing on the Antarctic plateau is rather unique as it is mostly located near the ground. The ice causes a strong thermal inversion that spans a few tens of meters. Above this height, the atmosphere is very stable and shielded from mixing with warm air by the circumpolar vortex. At Dome C, this boundary layer is between 20{\m} and 30{\m}, above which the seeing averages a quarter of an arcsecond \citep{Lawrence, Agabi2006}. A telescope placed at this height would therefore experience the world's best seeing conditions.

    \item {\bf Long nights:} High cadence observations are key for certain science cases such as the detection of electromagnetic emission from gravitational wave events. Following up exoplanets as well as fast transients such as kilonovae without being interrupted by the day/night cycle is crucial. This is only possible on the ground near the poles.

    \item {\bf Excellent weather}: The studies of \citet{Crouzet+2010, Crouzet+2018}, which analyzed data collected over four winter seasons with the ASTEP South refractor, demonstrated the great photometric stability of the Concordia site. The analysis showed that conditions were considered photometric approximately two-thirds of the time, with uninterrupted observing periods lasting up to 100 hours during the polar winter. The effective duty cycle for observations reached up to 92\%, emphasizing the station's unparalleled ability to support continuous monitoring. The study of weather conditions at different heights at Concordia over ten years (2010-2019) by \citet{Genthon+2021} shows that mean monthly temperatures on the ground are between $-60${\degC} and $-65${\degC} during the Antarctic night and temperature can reach as low as $-80${\degC}. Contrary to other locations in Antarctica and particularly on the coast, the wind conditions are very favorable, with mean monthly winds below 4.5\,m\,s$^{-1}$ on the ground, increasing to less than 7\,m\,s$^{-1}$ at 18{\m} elevation. These results highlight Concordia’s suitability for high-precision photometric campaigns, particularly those requiring extended and stable observing windows.

    \item {\bf Unique field of view:} The Antarctic location offers a unique opportunity to observe sources at high negative declinations, which are otherwise difficult to access with good conditions from most ground-based observatories. This is particularly useful for complementing space-based surveys of exoplanets, $\gamma$-ray sources, and other astrophysical objects, as well as ground-based experiments detecting gravitational waves or neutrinos. For instance, only a few percent of ASTEP’s long-period transits ($P<20${\days}) could have been observed elsewhere \citep{Dransfield2022}. Notably, Dome C's vantage point provides continuous visibility of the southern continuous viewing zones of {\it TESS}, {\it JWST}, and {\it Ariel}, as well as most of {\it PLATO}’s field of view. 

    \item {\bf Convenient logistics:} The Concordia Station is operated with people on-site throughout the year, the logistics being supported jointly by the French and Italian polar agencies, IPEV (Institut polaire français Paul-Émile Victor) and PNRA (Programma Nazionale di Ricerche in Antartide). Personnel and cargo reach the station either via overland traverse or air. The traverse, originating from the Dumont d’Urville station, covers approximately 1,200{\km}. These expeditions, often referred to as ``raids", typically take 7 to 12 days depending on weather conditions and are conducted three to four times per summer season. The convoys transport essential supplies like fuel and food. Personnel and lighter cargo can also arrive via aircraft, using a skiway at the station, with flights from Dumont d’Urville or other nearby Antarctic research hubs such as Mario Zucchelli Station. These logistical systems ensure a steady flow of resources and personnel to support scientific and operational needs year-round.

\end{itemize}

These conditions provide an ideal environment for wide-field observations in the infrared. The combination of high atmospheric transmission and low seeing directly translates to unprecedented sensitivity, which allows a small aperture telescope to compete with a much larger one located at temperate latitudes.

\subsection{Telescope design}\label{sec2.2}

At full scale, Cryoscope will be a 1.2{\m} telescope with a 50{\sqdeg} field-of-view in the $K_{dark}$ bandpass
while keeping thermal background well below the very dark $K$-band sky in the Antarctic. This requires a radically different approach to traditional NIR imagers which block telescope radiation by re-imaging the pupil onto a cold Lyot stop or by carefully baffling room-temperature telescopes (e.g., ESO's VISTA telescope). Both methods lead to higher background and yield much smaller fields-of-view than Cryoscope which cryogenically cools the entire optical path to reduce thermal background. The cryogenically cooled infrared detector is placed at prime focus with the evacuated telescope replacing the detector cryostat.

\begin{figure*}
	\includegraphics[width=\textwidth]{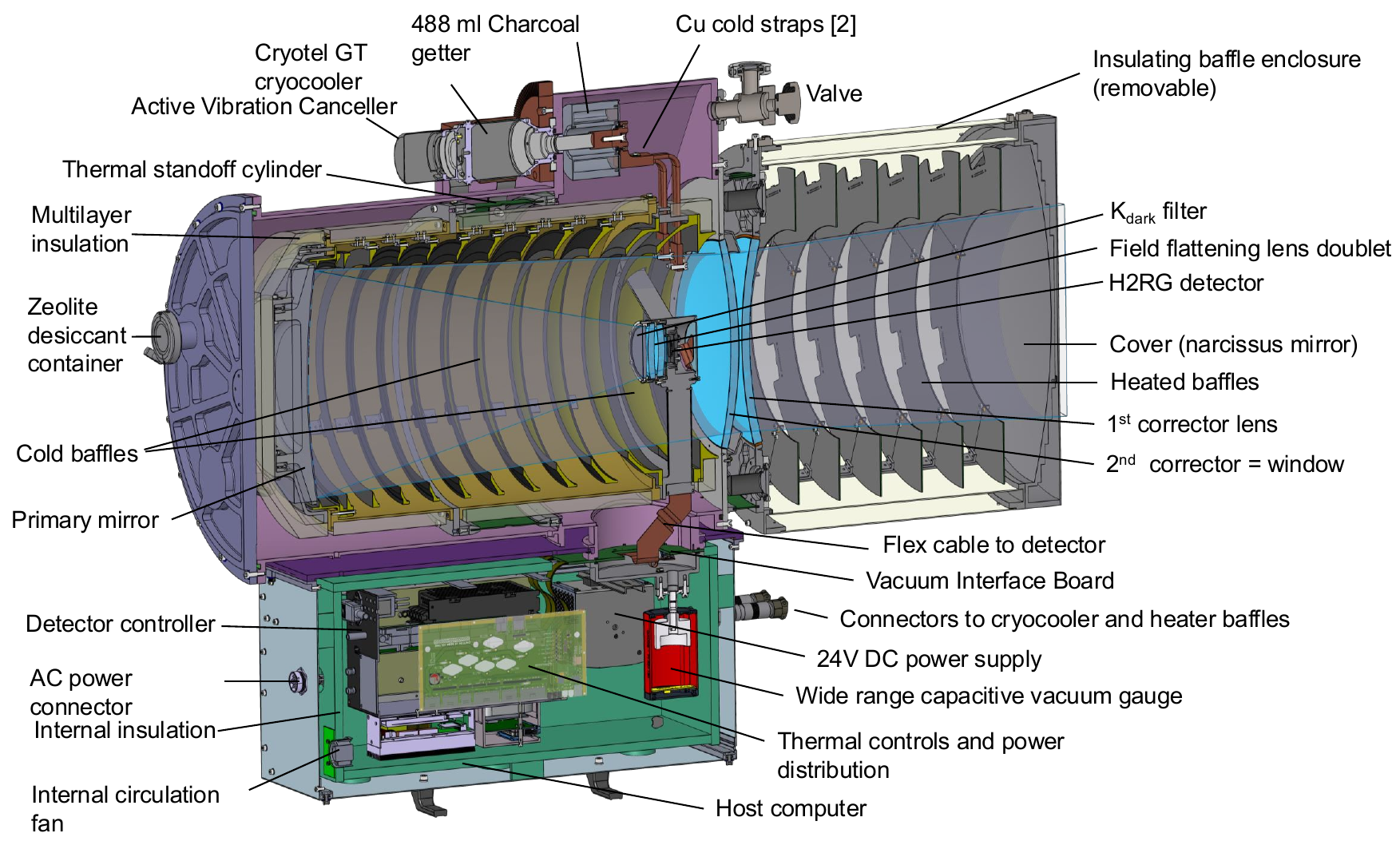}
    \caption{Cross-section of the Cryoscope Pathfinder, showing the major components of the design. The two corrector meniscus lenses at the entrance aperture correct for aberrations from the spherical primary mirror, with the 2nd corrector also serving as a window into an evacuated cryostat interior. Cold ellipsoidal mirror-baffles line the optics core and reflect the window's thermal emission back to the window to limit radiative cooling. Temperature controlled external radiative baffles limit the solid angle subtended to the first optical surface and maintain the first corrector lens at ambient temperature, preventing condensation from forming on the lens surface.}
    \label{fig:cross-section}
\end{figure*}

This deceptively simple configuration (Figure~\ref{fig:cross-section}) is made possible by a new optical design that employs a pair of meniscus lenses at the entrance to correct for aberrations created by the spherical primary mirror \citep{Fucik2022}. The 2nd meniscus is convex and thus able to support atmospheric pressure efficiently since it is in compression throughout its volume. This allows the telescope to be evacuated and cooled to suppress infrared emission.

\begin{figure*}
	\includegraphics[width=\textwidth]{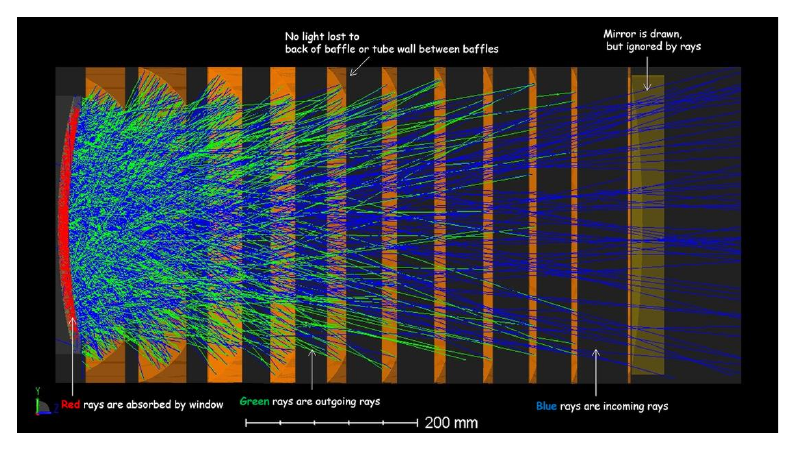}
    \caption{Ray tracing demonstrating Cryoscope's baffling scheme. The second corrector element (the window) is a source of thermal radiation (green rays). The ellipsoidal mirror baffles are designed to reflect the majority of the thermal radiation (blue) back to the window which is then absorbed by the lens' convex surface (red), preventing excessive radiative cooling.}
    \label{fig:ray}
\end{figure*}

The fused silica window is opaque in the mid-infrared where it radiates significant energy that would cool both meniscus lenses and result in condensation on the outer surface. To reduce this radiative cooling, $\sim 70\%$ of the window emission is reflected back to the window by elliptical mirror-baffles surrounding the beam (Figure~\ref{fig:ray}).
The remaining heat loss is replaced by warming the dry-air filled space between the two menisci, and by a small amount of radiation from the temperature controlled external baffles which intercept most of the solid angle seen by the first optical surface.

\subsection{The camera}
The baseline detector is a HELLSTAR detector, HgCdTe Extremely Large Layout Sensor Technology for Astrophysics Research, a device developed in the SATIN program by Rochester Institute of Technology (RIT) and Raytheon. The detector uses HgCdTe, deposited on silicon, as the light-sensitive layer bump-bonded to a read-out integrated circuit (ROIC) originally developed by Sensor Creations, Inc., now Attollo Engineering.

\begin{figure}
	\includegraphics[width=\columnwidth]{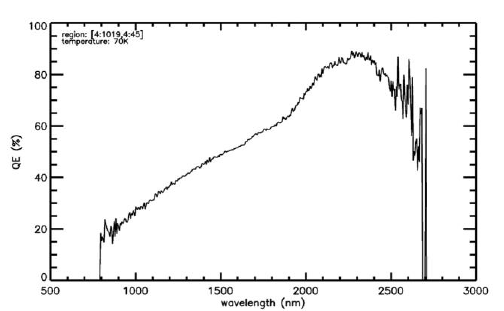}
    \caption{Quantum efficiency of a SATIN detector that uses HgCdTe deposited on silicon. SATIN is the infrared detector development program leading to HELLSTAR.}
    \label{fig:qe}
\end{figure}

\begin{figure*}
	\includegraphics[width=\textwidth]{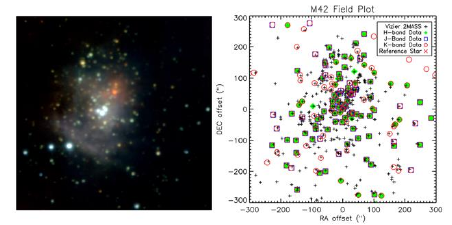}
    \caption{ {\it Left}: Color-composite {\it JHK} image of M42 obtained with a 1Kx1K version of a SATIN detector using HgCdTe deposited on silicon. The central four bright stars represent the Trapezium. The red object just above the center is the Becklin-Neugebaur object, demonstrating good sensitivity in the K-band.} {\it Right}: A plot of object locations from the data compared to those from 2MASS from \citet{Buntic2024}.
    \label{fig:M42}
\end{figure*}

The wavelength cutoff is tuned to 2.65{\um} in order to take advantage of the very low thermal background of the telescope and atmosphere. During the development projects leading to the final detector recipe, the team iterated the design across a range of pixel geometries and doping profiles \citep{Hanold2015,Figer2018,Figer2022,Buntic2024}. The final design delivers read noise of ~7 e$^-$ in Fowler-16 (25 e$^-$ CDS), QE near
90\% at the peak near 2.4{\um}, and dark current of 0.04 e$^-$/second/pixel. These performance values are all better than the requirements for Cryoscope, in part because the application is heavily background-limited. Figure~\ref{fig:qe} shows the QE, demonstrating the custom long-wavelength cutoff for Cryoscope. This device was also successfully demonstrated on-sky using the 24-inch telescope of the C.E.K. Mees Observatory in Bristol, New York. Figure~\ref{fig:M42} shows a color-composite $JHK$-band image of M42, clearly showing the Trapezium and Becklin–Neugebauer object, along with a chart showing corresponding stars from 2MASS data \citep{Buntic2024}. Table~\ref{tab:satin_hellstar} gives measured performance for detectors from the SATIN program along with goals for HELLSTAR.  

HELLSTAR has 4Kx6K pixels and could be arranged in a 9x6 mosaic of 54 detectors to create a 36Kx36K, or 1.36 gigapixel, focal plane. There are other possible detectors using Raytheon all-digital focal planes, such as the SB500 (4K$\times$4K$\times$18{\um}) and SB586 (4K$\times$4K$\times$10{\um}), the latter of which has been successfully acquired by the DKIST project.

\begin{table*}
	\centering
	\caption{Performance of SATIN detectors and goals for HELLSTAR detectors.}
	\label{tab:satin_hellstar}
	\begin{tabular}{c|c|c}
		 \textbf{Parameter} & \textbf{SATIN Performance} & \textbf{HELLSTAR Goals}\\
		\hline
		Format & 1K$\times$1K and 2K$\times$2K & 4K$\times$6K \\
            \hline
            Pixel size ($\,\mu$m) & 20 & 10 \\
            \hline
            Read noise (e$^-$, Fowler-16) & 10 & 10 \\
            \hline
            Wavelength range ($\,\mu$m) & 0.8-2.40 and 2.65 & 0.8-2.65\\
            \hline
            & 0.04 (70 K) & \\
            Mean dark current (e$^-$/s/pixel)\tablenotemark{a} & 0.07 (90 K) & 0.01 (70 K)\tablenotemark{b}  \\
            & 1.0 (110 K) & \\
            \hline
            Minimum readout time (s) & 0.3 & 1\\
            \hline
            Well size (e$^-$)  & 400,000 & 100,000\tablenotemark{b} \\  
            \hline
             & $>90\%$ in $K$-band & \\
            QE (w/ AR coating) & 50\% in $H$-band & $\geq90\%$ in $JHK$-band \\
            & $35\%$ in $J$-band & \\
            \hline
            Persistence (after 100\% saturation) & $< 0.2\%$ & $< 0.2\%$ \\
            \hline
	\end{tabular}
    \tablenotetext{a}{This result is measured after accounting for glow, as opposed to values in \citet{Figer2022}.}
    \tablenotetext{b}{This assumes geometric reduction of a factor $\sim 4$ due to smaller pixel size.}
\end{table*}

In preparation for its deployment, the fifth-scale prototype, Cryoscope Pathfinder, will first employ a legacy 2K$\times$2K$\times$18{\um} Teledyne H2RG detector array. We will then consider installing a larger device, such as HELLSTAR, or one based on the SB500 or SB586 ROIC loaned from the DKIST project.

\section{Prototype development}\label{sec:pathfinder}
Cryoscope's double meniscus corrector lenses enable its infrared camera to image a wide field-of-view under one of the darkest infrared sites on Earth at Dome C. In this section, we detail the status of the technology demonstrator for this new optical design.

\subsection{Description}
The Cryoscope Pathfinder is an f/2, 26{\cm} aperture, 16{\sqdeg} field-of-view imager which will retire technological risk for the future $1.2$-meter, 50{\sqdeg} field-of-view design. The field-of-view imaged by the Pathfinder at 7.1\,\,arcsec/pixel is limited by the available detector and not the optical design. The Pathfinder's modest 26{\cm} aperture is selected such that its legacy H2RG detector ($2048\times2048$ pixels, 18{\um}/pixel) and housing obstruct less than 15\% of the incident beam. The telescope's focal ratio approximates that planned for the meter-scale Cryoscope. 

\begin{table*}
	\begin{center}
	\caption{Cryoscope Pathfinder vs. Crysocope}
	\label{tab:proto_vs_full}
	\begin{tabular}{ccc}
		\hline
		 Parameter & Cryoscope Pathfinder & Cryoscope\\
		\hline
		Aperture (m) & 0.260 & 1.2 \\
            Focal length (m) & 0.525 & 2 \\
            Pixel size ($\,\mu$m) & 18 & 10 \\
            Focal plane area (Mpix) & 4.2 & 604 \\
            Plate scale (arcsec/pixel) & 7.07 & 1.03 \\
            Field-of-view (deg$^2$) & 16.2 & 49.6 \\ 
            Dark current (e$^-$/s) & 0.1 & 0.1 \\
            Read noise (e$^-$) & 15 & 15 \\
            Estimated Dome C sky background (e$^-$/s/pixel) & 210 & 140 \\
            Depth for 120 s visit{\tablenotemark{a}} (AB mag) & 18.0 & 21.9\\
            Depth for 1 hour visit{\tablenotemark{a}} (AB mag) & 20.2 & 24.1\\
            Areal speed (deg$^2$/hr) & 486 & $1.49\times10^3$\\
            Volumetric speed{\tablenotemark{b}} (Mpc$^3$/s) & $1.80\times10^2$ & $5.24\times10^4$ \\
		\hline
	\end{tabular}
    \tablenotetext{a}{Coadded stack of an 8-step dither sequence with 10\,s individual exposures of a point source, SNR=5}
    \tablenotetext{b}{120\,s visit (including estimated 40\,s overheads) maximizes survey speed for sources at $M_{AB} = -19$}
    \end{center}
\end{table*}

The Pathfinder's principal goal is to retire the technological risk inherent to the design of an Antarctic ultra-wide-field infrared surveyor detailed in \S\ref{sec2.2}. The Pathfinder will be the first infrared surveyor deployed at Dome C to measure the $K_{dark}$ sky brightness over a near-continuous winter observing season. To exploit the sensitivity gains driven by reduced sky background, the Pathfinder will also meet the requirement to minimize thermal background emission from the telescope itself to remain sky-background-limited.

\begin{figure*}
	\includegraphics[width=\textwidth]{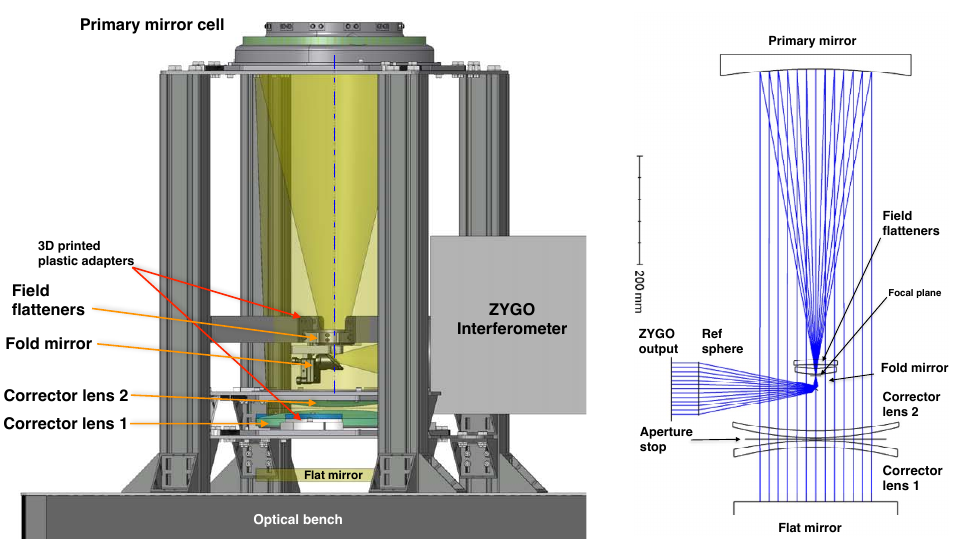}
    \caption{ {\it Left}: Mechanical layout of warm bench test setup of end-to-end system for interferometric wavefront measurement. {\it Right}: Optical layout of the test setup.}
    \label{fig:benchtest_setup}
\end{figure*}

The Cryoscope Pathfinder has been designed, fabricated, and assembled by the Caltech Optical Observatories. All lenses and the diamond-turned aluminum primary mirror have been tested as a full assembly at ambient temperature on the bench (Figure~\ref{fig:benchtest_setup}). The on-axis wavefront error of the double meniscus telescope has been measured using a Zygo interferometer by illuminating the telescope in double-pass \citep{Earley2024}. A reference sphere mounted at the exit port of the Zygo is used to create a f/2 beam matching the focal ratio of the telescope. The focused light from the Zygo enters the telescope via a small 45$^\circ$-fold mirror just behind the central obscuration where the telescope focal plane is located. The telescope converts the focused light into a collimated beam that is reflected by a 25\,cm diameter reference flat, so that it is returned to the Zygo along the same optical path. 

\begin{figure*}
	\includegraphics[width=\textwidth]{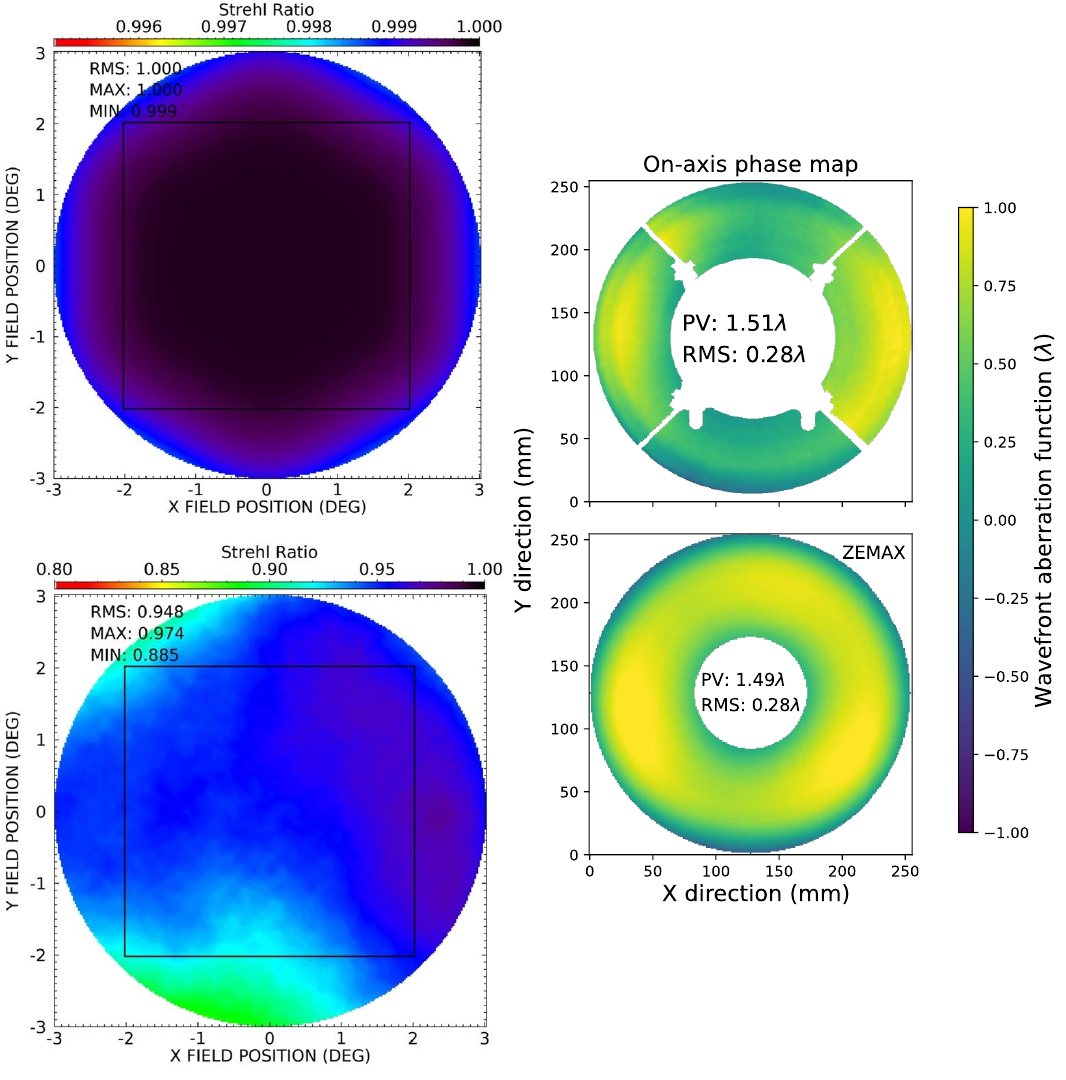}
    \caption{ {\it Top left}: The design delivers Strehl ratio $> 99.9\%$. The inscribed black square is the detector footprint. {\it Bottom left}: 95\% Strehl is predicted from surface measurements of optics as built. {\it Right:} On-axis measured wavefront error from \citet{Earley2024} in the top panel agrees with the design from as-built surface measurements, shown in bottom panel.}
    \label{fig:strehl_design}
\end{figure*}

The on-axis wavefront error in single-pass for the entire telescope is 0.28 wave rms at 632.8\,nm. This result agrees with the optical model for the test setup and confirms that the optical elements have been co-aligned to within $\pm 0.2$\,mm. Because the refractive elements are meniscus lenses with little optical power, the positional tolerances are relatively loose for an f/2 objective. The corrector lenses have been optimized for $K$-band imaging and suffer from sphero-chromatism at 632.8\,nm, which is well outside the operational passband. It is inferred from this that the in-band wavefront error is $< 0.03$ wave rms and thus will deliver diffraction-limited performance. The optical model predicts a Strehl ratio greater than 99\% over the full field for design errors alone and 95\% when wavefront errors measured for each surface are taken into account (Figure~\ref{fig:strehl_design}). The measured wavefront error indicates that alignment errors are sub-dominant and within the allocated tolerances, so the 95\% Strehl is anticipated at 2.4{\um} in-band center wavelength. The interferograms obtained in the warm bench tests at 633\,nm indicate that the integrated optical system will deliver the deeply diffraction limited performance (95$\%$ Strehl ratio) at 2.4{\um}, limited only by surface errors and not by design or alignment. 

\begin{figure*}
	\includegraphics[width=\textwidth]{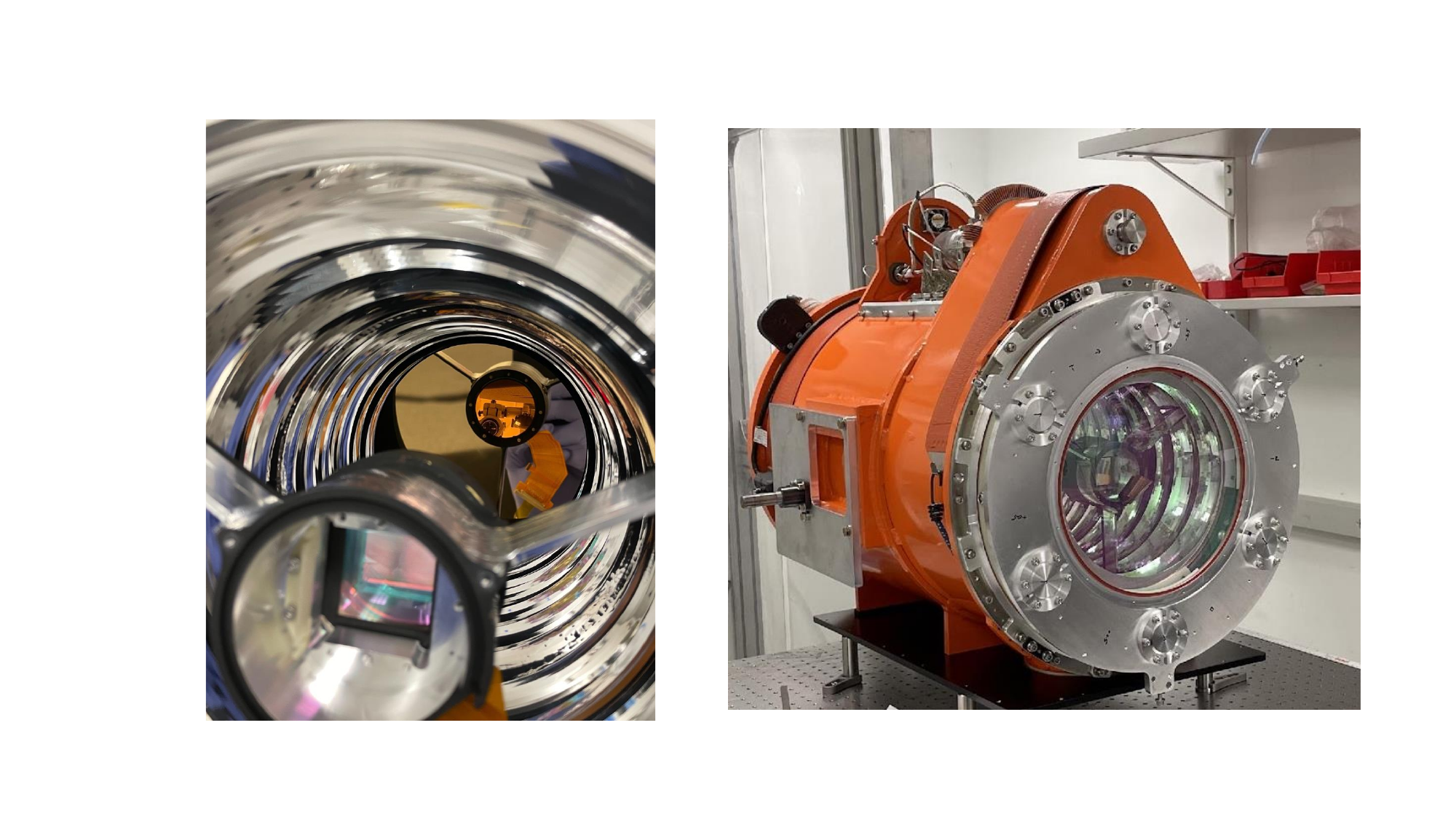}
    \caption{ {\it Left}: An interior look into the Cryoscope Pathfinder. The gold-plated aluminum spherical primary mirror lies in the background. The detector housing, prior to detector installation, is shown in the foreground. The housing supports the field flattener doublet optimized for the $K$-band. Lining the sides of the telescope are the ellipsoidal polished cold radiative baffles. {\it Right}: Full optical assembly and alignment has been completed on the bench. Not shown are the electronics box and warm radiative baffles exterior to the entrance aperture.}
    \label{fig:assembled}
\end{figure*}

The bench-tested optics have been installed and aligned in the cryostat (Figure~\ref{fig:assembled}). The $K_{dark}$ filter was manufactured by Asahi Spectra and has $> 99\%$ transmission in the passband with 50-50 points at 2.25{\um} and 2.5{\um}, and a center wavelength of 2.375{\um}, optimized for an angle of incidence of 9.65$^\circ$. All cryostat elements have been baked out, and zeolite dessicants ensuring dry air space between the corrector lenses have been installed. The convex corrector lens serving as the vacuum window has been shown to hold atmospheric pressure at environmental temperatures of $-85${\degC}, with no performance loss over a 2-week test. All seals in the system have successfully held vacuum over repeated temperature cycles from ambient lab temperature to $-40${\degC}. Testing to lower temperatures is ongoing.  Thermal interlocks, servos and telemetry are fully operational and have been validated during thermal cycles. The legacy H2RG detector has been installed at prime focus and is currently under test.  Sunpower's Cryotel GT Stirling cryocooler is employed to maintain the telescope and detector at operating temperatures of 100\,K. Initial testing revealed higher-than-expected final detector temperatures. As the cryocooler is evaluated, the detector readout and temperature-dependent dark current properties are being tested. Reducing detector temperature will reduce dark current to within required ranges based on estimated sky brightness levels. Higher-conductance cold thermal straps for the detector and optics core are currently being designed to further improve cryocooler performance. We first plan to implement a liquid nitrogen-based solution to complete lab testing and validation. We will then implement a cryocooler-based approach prior to deployment at Dome C. 

\subsection{Critical objectives}

Pre-shipment testing of the entire system is underway with the expected deployment date for Dome C in December 2026. The critical mission objectives for the Pathfinder are the following:

\begin{enumerate}
    \item \textbf{Validate image quality and focus control} of the fully assembled cryogenic, double meniscus optical design. The next steps are to show that the optical performance revealed in the lab can be replicated on-sky when cold. On-sky image quality will be verified at Caltech ahead of deployment to Dome C for the 2026 season. 
    Final validation will require that the optical performance not be degraded by atmospheric pressure on the window or cryogenic cooling of the primary mirror and field flatteners. Cold tests will demonstrate focus control and stability, accomplished by adjusting the operating temperature within the cryostat. Temperature variations of 1$^\circ$\,C correspond to 10{\um} of defocus. Shims are used to set the nominal focus position at a specific operating temperature. The adjustment range via thermal control will be validated, assuring that once shimmed correctly, no further adjustments will be required.
        
    \item \textbf{Suppress the telescope's thermal self-emission below Dome C sky background levels.} The novel double meniscus optical design enables one corrector lens to serve as a vacuum window so the telescope's interior can be held at cryogenic temperatures. Suppression of instrumental thermal emission ensures that observations are sky-background-limited without limiting the field-of-view. Lab tests will demonstrate this concept of thermal background suppression by first capping the entrance aperture with a reflective mirror and measuring the background signal as a function of mirror temperature. Extrapolating the background signal at the mirror's coldest temperature to the background expected in the absence of the mirror's thermal emission, we will quantify the signal that arises exclusively from thermal self-emission. Following deployment at Dome C, the background from the telescope will be measured again using the reflective cap, but this time at much lower ambient operating temperature.  After the emission by the telescope has been shown to be subdominant, Pathfinder will measure the $K_{dark}$ sky brightness throughout the observing season.
    
    \item \textbf{Avoid condensation on the corrector lenses.} Ellipsoidal cold mirror baffles in the interior of the cryostat (left panel, Figure~\ref{fig:assembled}) are designed to reflect $\sim 70\%$ of the window's thermal emission back to the window (Figure~\ref{fig:ray}). 
    The remaining heat loss is replaced by warming the dry-air filled space between the two menisci, and by a small amount of radiation from temperature controlled external baffles (not shown in the right panel of Figure~\ref{fig:assembled}). The effect of condensation prevention strategies on the thermal background will be assessed. Supplemental dry air injected in front of the first corrector lens will be available at Dome C to further discourage condensation.   

\end{enumerate}

\subsection{Implementation at Concordia}

\begin{figure*}[tb!]
  \includegraphics[width=\textwidth]{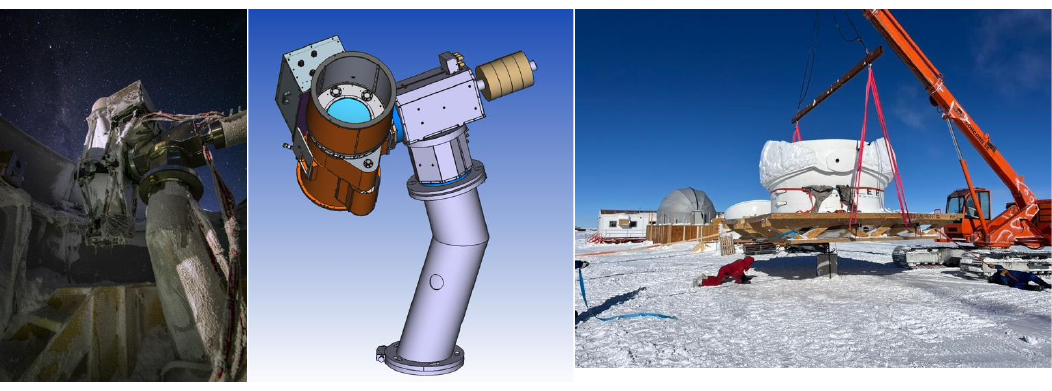}
  \caption{ {\it Left:} ASTEP telescope and mount in operation (Photo: V. Deloupy, 14 June 2024). {\it Middle:} CAD design of Cryoscope Pathfinder on the twin ASTEP direct drive mount. The Pathfinder will be installed on a mount identical to the one used by ASTEP. {\it Right:} Installation by the IPEV and PNRA teams at the Concordia Station of the pillar and dome to be used by Cryoscope Pathfinder, at latitude 75.100$^\circ$S, longitude 123.319$^\circ$E, and an elevation of 3233\,m. (Photo: A. Agabi, 21 Nov. 2024).}\label{fig:Concordia}
\end{figure*}

The proposal to install and operate the Cryoscope Pathfinder at the Concordia base was approved by the French polar institute IPEV in February 2024. The Pathfinder will operate in a way similar to that of ASTEP, with the possibility of combining scientific goals, as described in \S\ref{sec:science}. 

The Pathfinder will be installed on a direct-drive mount identical to the one used by the ASTEP telescope  (Figure~\ref{fig:Concordia}). This mount has been designed at OCA to cope with the extremely cold temperatures of Antarctica and has operated continuously with no failure since March 2024.

In November and December 2024, the IPEV and PNRA teams at Concordia installed a pillar, dome, and conduits designed to house power and data cables, as well as the necessary infrastructure to operate Cryoscope Pathfinder in December 2026.

Concordia Station is connected to the Internet via a satellite link. Although the available bandwidth is limited, it provides a data transmission capacity of 20 GB/day for Cryoscope Pathfinder. Until bandwidth capacity is increased, data processing will be performed using pipelines deployed on an on-site server, with results subsequently transmitted.
The station's Internet connection enables remote control of the telescope's operation and data acquisition.

\section{Science cases}\label{sec:science}

Next, we describe nine science cases that drive the Cryoscope design. While this list is not exhaustive, it illustrates the breadth in astronomy enabled by a wide-field infrared surveyor. 

\subsection{Exoplanets}

Exoplanetology is a rapidly evolving field, with unexpected discoveries occurring routinely. This science case benefits from a robust fleet of operational and planned spacecraft. For the detection of new exoplanets, NASA's {\it TESS} \citep[Transiting Exoplanet Survey Satellite;][]{Ricker2014} has produced over one thousand candidates in the past two years, outpacing, and also driving an extensive follow-up work involving additional photometry, radial velocity measurements, and direct imaging campaigns to confirm and characterize these discoveries. In parallel, the focus of the field has increasingly shifted towards atmospheric characterization. Pioneered by the {\it Hubble} Space Telescope, this effort has entered a transformative phase with the advent of the James Webb Space Telescope ( {\it JWST}), launched in 2021, which provides unprecedented high signal-to-noise spectra for robust atmospheric studies \citep[e.g.,][]{Ahrer2023}.

Looking ahead, major milestones include the launch of {\it PLATO} \citep[][]{Rauer2014} in 2026, the first light of the European Extremely Large Telescope (E-ELT), and ESA's {\it Ariel} mission \citep{Tinetti2018}, both slated for 2029. Together, these facilities will advance the field through a range of techniques, including low- and medium-resolution transmission and emission spectroscopy from space, as well as high-resolution spectroscopy with the E-ELT \citep{Deming2017}.

The following paragraphs highlight the main science themes driving exoplanet research and detail how extending observations into the infrared with Cryoscope will amplify these efforts, leveraging not only the enhanced sensitivity of the proposed telescope but also the unique advantages offered by the long, stable conditions of Antarctic winters at Dome C.

{\bf How do planetary systems form?}
Despite decades of research, many questions on how planetary systems form remain unanswered. While core accretion \citep{Pollack1996} must account for the formation of most known giant planets, direct formation by gravitational instability may be needed for some gas giant planets, particularly ones at large orbital distances \citep[e.g.,][]{Cadman+2021}. Open questions also concern how giant planet cores are assembled, the roles of planetesimals versus pebble accretion, and the impact of photoevaporation on planetary compositions \citep{Lambrechts+Johansen2012, Helled+Stevenson2017, Guillot+Hueso2006, Morbidelli+2024}. More broadly, the relationship between stellar mass and the architectures of planetary systems remains unclear \citep[e.g.,][]{Owen2013, Owen2017, Eylen2018, Venturini2020}. This is exemplified by the ongoing debate on whether small exoplanets exhibit a radius gap or a density gap \citep{Luque2022}. Progress on these topics requires advancements in three areas: (i) detecting and characterizing more long-period planets, which are rarer but provide essential statistical insights \citep[e.g.,][]{Thorngren2018,Guillot2022}; (ii) studying planets around diverse stellar types, particularly M dwarfs, which require infrared observations; and (iii) characterizing entire planetary systems, especially those with interacting planets, which provide critical constraints on formation scenarios.

{\bf Looking for temperate planets}:
Discovering long-period planets, particularly those in temperate zones conducive to habitability, is a critical goal for exoplanet science. These planets, often located far from their stars, provide essential clues about the diversity of planetary systems and their potential for hosting life. However, their detection remains challenging due to their faint signals and the extended time required to constrain their orbits. Ground-based observatories, such as the ASTEP (Antarctic Search for Transiting Planets) program \citep{Guillot+2015,Crouzet2020,Schmider2022,Dransfield2022}, have demonstrated the potential of Antarctica's unique conditions -- long, uninterrupted nights, and excellent visibility of the continuous viewing zones of key space missions like {\it TESS}, {\it JWST}, and soon {\it Ariel}.

The upcoming {\it PLATO} mission, set to launch in 2026, will revolutionize the search for long-period planets by targeting bright stars and extending sensitivity to planets on longer orbits. Notably, {\it PLATO}'s field-of-view has been strategically selected to include regions with high visibility from Antarctica's Concordia Station, enhancing the potential for ground-based synergy with missions like ASTEP \citep{Nascimbeni2022}. Vetting the candidates identified by {\it PLATO} is vital to confirm their planetary nature and refine their properties. Simultaneous observations at various wavelengths, particularly in the infrared, could play a pivotal role in this follow-up work, enabling the efficient detection of interlopers such as background eclipsing binaries.

\begin{figure}
  \includegraphics[width=\columnwidth]{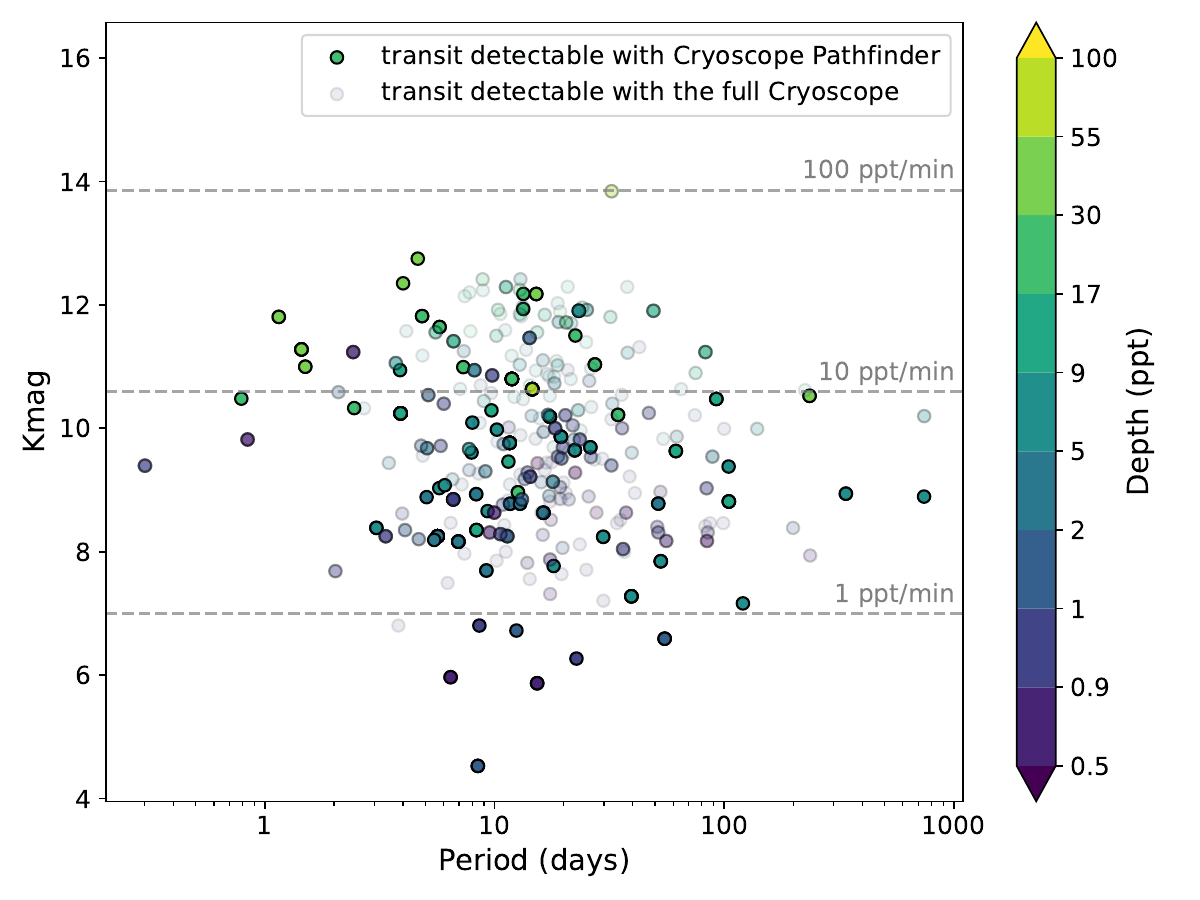}
  \caption{ {\it TESS} planetary candidates observed by ASTEP from 2022 to 2024, plotted by $K$-band Vega magnitude versus orbital period. ASTEP’s focus on long-period planets has led to the discovery of several transiting planets with orbital periods exceeding 100 days. All these planetary transits would be observable with high signal-to-noise ratio using Cryoscope, and over 20\% of them (highlighted) would be detectable with Cryoscope Pathfinder. The color bar illustrates the transit depth produced by an exoplanet, in ppt (parts per thousand in relative photometry). \label{fig:K_vs_P_ASTEP2022-2024}}
\end{figure}

Figure~\ref{fig:K_vs_P_ASTEP2022-2024} shows that ASTEP and Cryoscope Pathfinder can be usefully combined to characterize transiting temperate exoplanets with orbital periods beyond tens of days. In addition, Cryoscope Pathfinder has the possibility to detect transits of planets around M-dwarfs currently inaccessible to ASTEP.   

{\bf Looking for terrestrial planets (around M dwarfs)}:
Arguably, the leading science case in exoplanetology is the identification of temperate, rocky terrestrial planets in transit, which would enable explorations of their atmospheric properties and evidence for biology beyond Earth. Planets of this type are currently easiest to detect if they orbit ultra-cool 
dwarfs, stars with masses $< 0.15~\rm  M_\odot$. Their reduced size and luminosities enable planetary signatures to be $100\times$ deeper and occur $50\times$ faster than for similar planets orbiting Sun-like stars \citep[e.g.,][]{Triaud2021}. Several successes have been registered by telescopes of modest size such as TRAPPIST and SPECULOOS that produced the spectacular TRAPPIST-1 system, host to seven rocky temperate planets \citep{Gillon2016,Gillon2017}. The few known habitable-zone rocky worlds orbiting ultra-cool dwarfs imply an occurrence rate of order 30 to $50\%$ \citep{He2017,Triaud2021}. Since ultra-cool dwarfs are $3\times$ more frequent than Sun-like stars, there could be $10\times$ as many habitable-zone rocky worlds orbiting ultra-cool dwarfs than around Sun-like stars in our Galaxy. These planets are thus essential for studying the most frequent outcome of terrestrial planet formation and detecting or placing upper limits on the frequency of biology in the Cosmos \citep{Triaud2021}. Detecting such planets requires long photometric timeseries \citep[$\geq 100 ~\rm hrs$;][]{Delrez2018,Sebastian2021}. The detection of such systems is also an opportunity to study the outcome of planet formation around stars with a mass $10\times$ smaller than the Sun's and to measure how physical laws underpinning planet formation scale with stellar mass. Ultra-cool dwarfs have a peak luminosity in the near-infrared. Many of them are also variable, and stellar variability reduces in the infrared.

{\bf Systems with TTVs}:
Planet-planet gravitational interactions produce variations in the mid-time of transits \citep{Agol2005}. Planets in mean-motion resonances are particularly sensitive to such interactions. These TTVs (Transit Timing Variations) are both an issue and a boon. They are an issue because it is no longer simple to predict the next transit time since the three-body problem does not have an analytical solution. For predictions, we must instead rely on computationally intensive N-body integrations matched to the data, which complicates reliably triggering large facilities that require observing during transit for exo-atmospheric studies (e.g. {\it JWST}). However, these very gravitational interactions measure bulk densities in absolute terms \citep{Wu2013}, thus providing the most direct access to geophysical information about exoplanets' interiors. ASTEP is particularly attuned to investigate systems with TTVs. Typically, a planet with an orbital period $> 15 ~\rm days$ only has one transit observable, in full, per calendar year from a given observatory. Due to the near continuous Antarctic winter night, ASTEP can capture multiple transits in one season for objects that are the most interesting for atmospheric studies \citep[since TTVs can give unusually precise bulk densities; e.g., ][on TRAPPIST-1]{Agol2021}, but which conversely are the hardest to schedule. Cryoscope will build on ASTEP's heritage in this area.

{\bf Characterizing Atmospheric Dynamics}:
Atmospheric dynamics of exoplanets are probed through techniques like transmission and emission spectroscopy, as well as phase curve observations. Transmission spectroscopy, performed during transits, reveals the atmospheric composition and structure, while emission spectroscopy, measured during secondary eclipses, provides insights into dayside temperatures and pressure-temperature profiles. Phase curves, which track the waxing and waning of the planet's dayside emission, are particularly valuable for studying heat redistribution and night-side temperatures. These measurements are the primary tool for characterizing atmospheres of non-transiting planets in the absence of direct imaging \citep{Brogi2012}.

Despite their scientific richness, phase curve observations are rare due to their demanding time requirements, often exceeding one full orbital period. Spacecraft like {\it Hubble} and {\it JWST} focus on short-period planets due to resource constraints, and only one ground-based phase curve measurement has been reported -- a three-week observation of WASP-19b by ASTEP from Antarctica \citep{Abe2013}. Moving to the infrared enhances emission and phase curve spectroscopy by exploiting the favorable planet-to-star flux ratio, amplifying the atmospheric signal and unlocking details about thermal dynamics as the planet orbits its star.

{\bf Cryoscope's Assets:}
While Cryoscope's modest aperture -- $0.26~\rm m$ for the Pathfinder and $1.2~\rm m$ for the full Cryoscope -- may seem small compared to telescopes like {\it JWST} ($6.5~\rm m$) or the E-ELT ($39~\rm m$), size is not everything. For exoplanet studies, time is a more critical factor than sheer photon collection. Long-duration observations, such as covering entire transits or phase curves, are essential, and polar sites like Dome C provide a unique advantage with nearly uninterrupted nights. ASTEP's success in obtaining long photometric sequences (e.g., 48 hours) from this location demonstrates how smaller telescopes can complement larger ones, particularly for observing planets that are hard to observe from mid-latitudes; only 6\% of ASTEP's {\it TESS} candidates are fully observable from such facilities \citep{Dransfield2022}.

With Cryoscope, we aim to improve sensitivity in the near-infrared and reduce telluric noise. Long-period transiting planets, typically observed by {\it TESS}, require continuous monitoring to maintain accurate ephemerides, which are crucial for atmospheric studies. While spacecraft like {\it JWST} and the E-ELT rely on precise transit timings, their effectiveness is limited by shifting ephemerides. Ground-based observatories, like Cryoscope, provide a cost-effective solution to update these timings and maximize the use of facilities in continuous viewing zones. For example, from Dome C, we can observe key transiting planets year-round, while other southern observatories are constrained by weather and limited visibility, offering a duty cycle of less than 30\% during peak visibility.

Modest telescopes have already shown their potential. SPECULOOS excels in studying ultra-cool dwarfs due to its combination of aperture size and red sensitivity \citep{Delrez2022}, and UltraCAM is optimal for white dwarfs thanks to its fast readout times \citep{Dhillon2007}. By focusing on the $K_{dark}$ band, Cryoscope aims to make groundbreaking observations, with the Pathfinder playing a critical role in ensuring we meet these ambitious goals.

\subsection{Neutron Star Mergers}

On August 17, 2017, the groundbreaking discovery of both gravitational waves and electromagnetic radiation from a binary neutron star (NS) merger marked the dawn of a new era in multi-messenger astrophysics  \citep{GW170817}. GW170817 lit up the entire electromagnetic spectrum, spanning $\gamma$-rays to radio, and yielded a scientific bonanza in fields as wide-ranging as strong field gravity, nucleosynthesis, extreme states of nuclear matter, the astrophysics of relativistic explosions, and even cosmology \citep{MMA}.
For the first time, we saw direct evidence of r-process nucleosynthesis, the process by which half of the elements in the periodic table heavier than iron are synthesized \citep{1976ApJ...210..549L}. Heavy line blanketing from the large density of bound-bound transitions renders the opacity of r-process rich matter much higher than the conventional iron peak elements, shifting the emergent spectrum of the ``kilonova" out of the optical bands and into the infrared \citep{Metzger:2010MNRAS.406.2650M, barnes:2013ApJ...775...18B,kasen:2013ApJ...774...25K}. 

The infrared data are therefore key to understanding the nucleosynthesis of heavy elements by the r-process. Analysis of the infrared photometric evolution and the vivid broad features in the infrared spectroscopic sequence showed that at least 0.05 solar masses of heavy elements were synthesized by GW170817 \citep{Coulter2017,Drout2017,Evans2017,Kasliwal2017b, Smartt2017, SoHo2017,Cowperthwaite2017,Arcavi2018}.
Combining the observed ejecta mass with rate estimates indeed gives numbers in the ballpark of the observed r-process abundances in the solar neighborhood. However, it is much debated whether the distribution of elemental abundances resembles the solar distribution. Evidence for the synthesis of the heaviest elements comes from late-time mid-infrared studies \citep{Kasliwal2019b}.  

This first discovery opens up many questions for future discoveries to answer. Was GW170817 a representative NS-NS merger? Are NS-NS mergers the only sites of r-process nucleosynthesis? Do NS-NS mergers produce heavy elements in the same relative ratio as seen in the solar neighborhood? Are the heaviest elements in the third r-process peak, such as gold and platinum, synthesized in NS-NS mergers? Which elements are synthesized when a neutron star merges with a black hole (BH)? 

To answer these questions by localizing additional gravitational wave events, many wide-field optical cameras are being developed at all scales, including the Rubin Observatory. However, possibly none of the NS-BH mergers and only a subset of NS-NS mergers may radiate in the optical \citep{PhysRevD.88.041503,Kasen2017,Bulla2023}. Specifically, it is predicted that the optical emission is limited to polar viewing angles in NS-NS mergers where the velocities are relatively high or the opacities are relatively low due to neutrino irradiation \citep{2014ApJ...780...31T}. If the merger remnant collapses promptly to form a BH, the optical emission is suppressed \citep{2018PhRvD..98h1501F}. Despite extensive searches (e.g., with GROWTH~\citep{Kasliwal2020}, GRANDMA~\citep{2020MNRAS.497.5518A}, GOTO \citep{2020MNRAS.497..726G}, J-GEM collaboration \citep{2021PTEP.2021eA104S}), no kilonova was identified during the LIGO-Virgo third observing run (April 1, 2019 to March 27, 2020) and first part of the fourth observing run \citep[][Pillas et al., in prep]{ahumada:2024PASP..136k4201A}. The small number of BNS mergers in the third and fourth observing runs have lowered their volumetric rate estimate. Furthermore, no additional kilonova detection since GW170817 also constrains the optical luminosity function and we can now infer that the majority of kilonovae are not as intrinsically luminous in the optical as GW170817 \citep{Kasliwal2020, ahumada:2024PASP..136k4201A}.

On the other hand, all NS-NS and NS-BH merger models predict that bright infrared emission from radioactive decay of heavy elements is ubiquitous and much less dependent on mass ratio, remnant lifetime, viewing angle, opacity distribution and velocity distribution \citep{Kasen2017,Bulla2023,Kruger2020}. Moreover, higher ejecta mass may imply that NS-BH mergers are especially luminous in the IR. For instance, Pillas et al. (2025) and \citet{Kawaguchi_2020} show that the $K$-band distribution of the peak luminosity time for kilonovae associated with NS-BH mergers is broader and less sharply peaked compared to the $g, r, i$ band distributions, which peak around one day post-merger. This finding emphasizes the need for more intensive follow-up at various times to capture the peak accurately. Thus, there is a critical need for a wide-field infrared transient survey that hunts for the elusive electromagnetic counterparts to gravitational waves \citep[e.g.,][]{Frostig2022}.

\begin{figure}
    \includegraphics[width=\columnwidth]{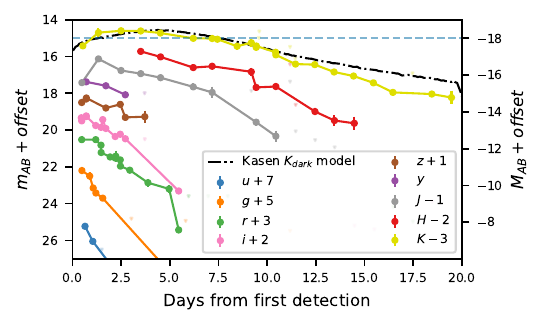}
    \caption{Multi-band photometry of GW170817/AT2017gfo from \citet{Kasliwal2017b}. GW170817/AT2017gfo faded rapidly in the blue bands but was long-lived in the infrared. No $K_{dark}$ data was obtained. Extrapolating from $K_s$ using Cryoscope's $K_{dark}$ filter and the best-fit kilonova model from \citet{Kasen2017}, the kilonova was brighter than $-15$\,\,mag for over 5{\days}.}\label{fig:knlc}
\end{figure}

\begin{figure}
    \includegraphics[width=\columnwidth]{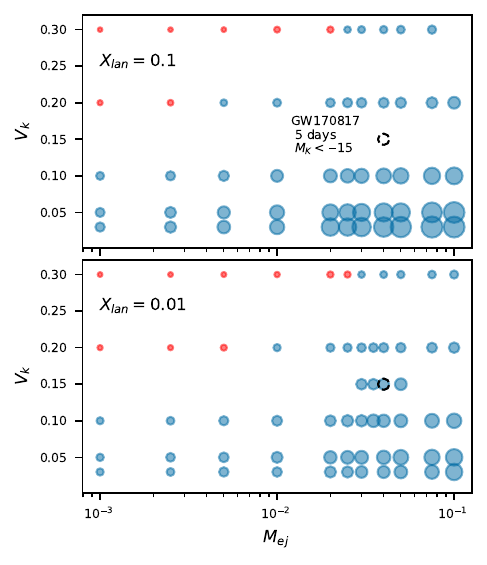}
    \caption{Model grids for kilonovae ejecta mass, velocity and lanthanide fraction \citep{Kasen2017}. Cryoscope will be sensitive to detecting kilonovae spanning a wide range of parameter space. All blue dots indicate the kilonova is brighter than $-15$\,\,mag for more than two days (size of the symbol proportional to duration). NS-BH mergers are expected to have a larger lanthanide fraction than NS-NS mergers like GW170817.}\label{fig:knmodels}
\end{figure}

\begin{figure}
    \includegraphics[width=\columnwidth]{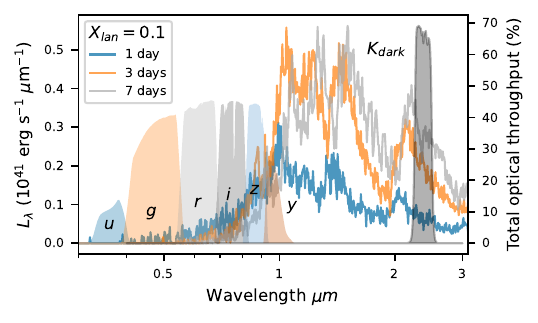}
    \caption{Three model kilonovae spectra shown at three snapshots in time after the explosion. The expected total optical throughput (right hand vertical axis) in the six Rubin filters is also shown, for comparison to that of Cryoscope's $K_{dark}$.}\label{fig:knspec}
\end{figure}

Examining GW170817, we find that the light curve evolution was fastest in the blue filters and slowest in the red filters (Figure~\ref{fig:knlc}). Specifically, in $K_{dark}$, the kilonova stays brighter than $-15$\,\,mag for over 5{\days}. Furthermore, if we examine kilonova model grids which depend on ejecta mass, ejecta velocity and lanthanide fraction, we find that a depth of $-15$\,\,mag would be sensitive to most of the model phase space (Figure~\ref{fig:knmodels}). Enhanced optical efficiency in $K_{dark}$ at Dome C guarantees that Cryoscope is well-suited to detect and monitor the luminosity evolution of the most lanthanide-rich merger scenarios (Figure~\ref{fig:knspec}).

\begin{figure}
\includegraphics[width=\columnwidth]{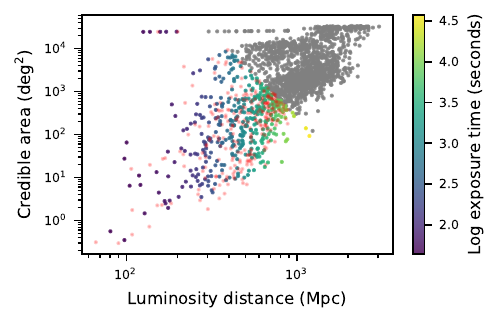}
\caption{Simulation of neutron star mergers for the fifth gravitational wave observing run planned at the end of this decade.  Color-coding indicates the exposure time needed for Cryoscope to map the 90\% GW localization area to find a kilonova brighter than $-15$\,\,mag. Gray points indicate events that are undetectable, requiring more than 10$^5$ seconds to tile the region to required depths. Light red points indicate events in the Northern hemisphere that are not accessible from the Antarctic location. Note that Cryoscope can detect all NS mergers within 300 Mpc irrespective of the localization and even detect well-localized NS mergers ($<300${\sqdeg} credible regions) out to 1 Gpc.}\label{fig:O5}
\end{figure}

Setting this $-15$\,\,mag depth as a threshold, we undertake a simulation of the fifth LIGO-Virgo-KAGRA gravitational wave observing run planned for 2028--2030. Using the latest estimates of the sensitivities of the GW interferometers \citep{Kiendrebeogo2023} and the latest release of the observing scenarios, we look at the detectability of kilonovae with Cryoscope (Figure~\ref{fig:O5}).  We find that Cryoscope is sensitive to all NS-NS and NS-BH events out to 300\,Mpc irrespective of localization of thousands of square degrees. Furthermore, Cryoscope is sensitive to events out to 1\,Gpc for localizations better than 100{\sqdeg}. 

By design, Cryoscope will be an immensely powerful kilonova discovery machine. Cryoscope will firmly and quantitatively answer the question of the mass of heavy elements that are synthesized by binary neutron star and neutron star-black hole mergers.  

\subsection{Galactic Stellar Science}

Cryoscope and its fifth-scale Pathfinder will enable many facets of Galactic stellar science. Here, we present a few examples. 

\begin{figure}
	\includegraphics[width=\columnwidth]{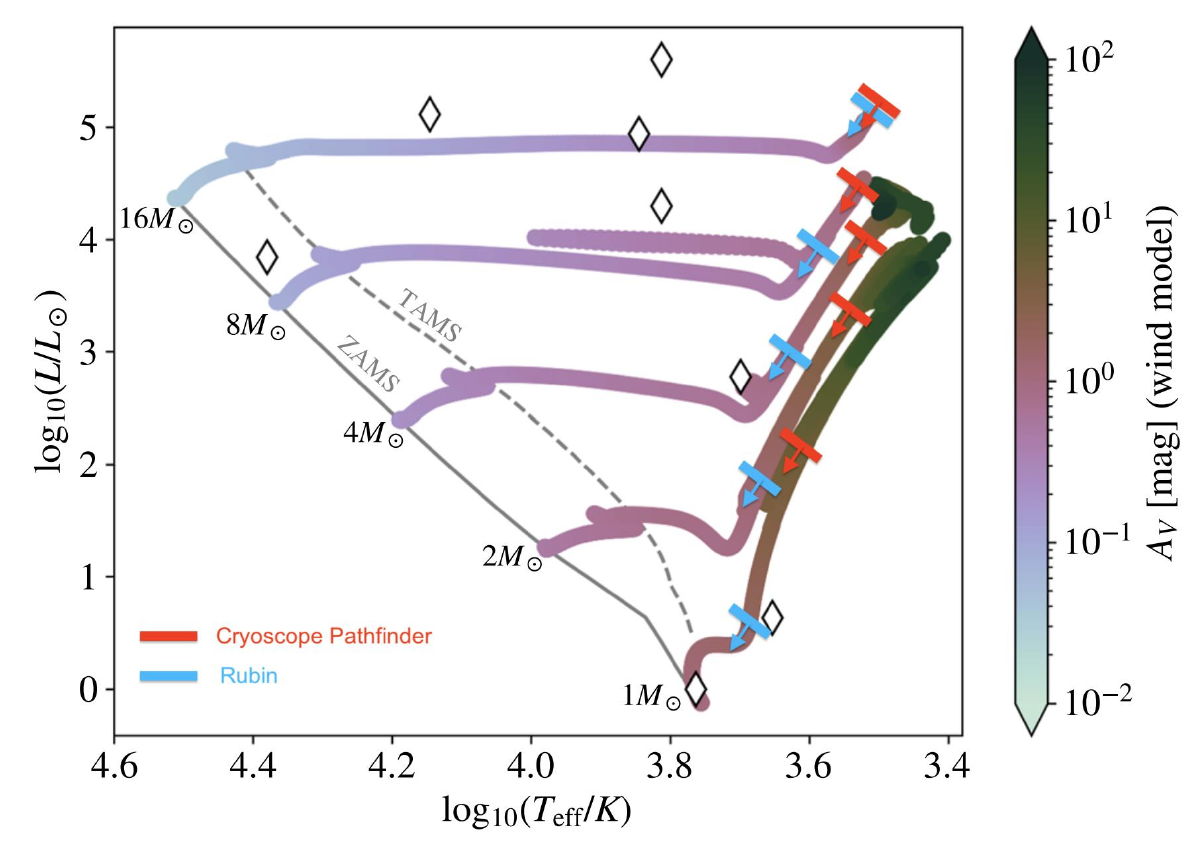}
    \caption{Evolutionary tracks \citep{MacLeod2022} for merging binaries with different primary masses color-coded to denote the expected extinction (see color-bar on the right). Upper limits are shown for the merger evolutionary stages accessible to the Rubin Observatory $r$-band (light blue) and for Cryoscope Pathfinder (red) for a Galactic merger at 10 kpc. The Pathfinder sensitivity rivals that of Rubin due to the $K_{dark}$ bandpass being much more immune to extinction than the visible bandpass. The expected rate of discovery is about one per year in the Milky Way and up to a few per year in the nearby galaxies. Even a single such discovery by Cryoscope Pathfinder would likely lead to a high-impact scientific result as it probes a hitherto unexplored region of parameter space in stellar merger physics. If the Pathfinder does not see such extremely extincted mergers, it would severely constrain theoretical models.}
    \label{fig:StellarMergers}
\end{figure}

{\em Stellar mergers}, particularly close binary systems undergoing mass transfer and subsequent merger, have a profound impact on stellar mass function \citep{Wang2022}. These mergers give rise to short period binary systems \citep{Tauris2006} and may be observed as luminous outbursts of the primary star followed by the slow ejection of the cool outer envelope and the formation of dust \citep{Ivanova2013}. Understanding them is crucial for studying stellar populations, but their identification is challenging due to their predominantly infrared emission as shown in Figure~\ref{fig:StellarMergers} \citep{MacLeod2022}. Cryoscope Pathfinder aims to overcome this challenge by efficiently detecting stellar mergers through a combination of short-lived optical outbursts and long-lived infrared emission. It provides valuable insights into the demographics of these mergers, including the elusive lowest-mass mergers, which are difficult to identify through optical surveys.

{\em Colliding-wind binaries (CWB)}: Dust is essential for chemical enrichment, planet formation, and life. Fundamental questions remain to be answered regarding the dominant channel for dust formation across cosmic time. CWBs have been identified as good dust producers that may explain quantities in the early universe \citep{WilliamsHucht1987, Lau2020}. Observational challenges in our galaxy are rooted in detecting these dust production episodes in the optical. However, Cryoscope Pathfinder, with its IR sensitivity, would be an ideal tool for discovering IR outbursts in CWB systems, facilitating subsequent mid-IR observations together with a program using {\it JWST}.

{\em Ultracompact binaries (UCBs)} are important for characterizing accretion physics in a unique regime. UCBs with white dwarf components emit both electromagnetic and gravitational radiation, enabling a better characterization of the source properties. Understanding UCB formation is crucial for studying Type Ia supernova progenitors and binary evolution. Detecting UCBs with cold companions (as a subclass called AM CVns $\sim 10$), is challenging. Cryoscope Pathfinder's $K$-band search can overcome the shortcomings of visible-wavelength surveys and reveal tens of non-outbursting AM CVns within a year. This will enhance our understanding of nearby gravitational wave emitters to prepare for the {\it LISA} space mission \citep{Breivik2018, vanRoestel2021}.

{\em Young stellar objects}: Star formation occurs through the collapse of molecular cloud cores \citep[see][for a review]{Armitage2019}. This process involves mass transfer from the cloud to the central protostar, marked by notable accretion outbursts. However, a comprehensive understanding of the amplitudes and timescales of these outbursts is still lacking. The challenge lies in the fact that most of the process is invisible in optical bands, where current time domain surveys excel in detecting outbursts. In contrast, infrared bands offer an opportunity to characterize these outbursts throughout all phases of the accretion process \citep{Armitage2019,Fischer2023}. The Cryoscope Pathfinder's time-domain survey of the Galactic plane will systematically uncover hundreds of young star outbursts in the visible Galactic disk.

{\em Novae}: Classical novae constitute some of the oldest known explosive transients, caused by thermonuclear ignition on the surface of an accreting white dwarf \citep{Starrfield2016}. In a small fraction ($\sim 10\%$) of cases, classical novae form dust shortly (weeks to months) after outburst, as characterized by dramatic dips in their optical light curves \citep{Williams2013}. Recent work suggests an intricate link between dust formation and their identification as luminous $\gamma$-ray transients \citep{Chomiuk2021}. A theoretical prediction that dust formation should be ubiquitous in novae, but observable as optical dips only along lines of sight near the binary orbital plane \citep{Derdzinski2017} results in a testable prediction. As some of the brightest transients in the infrared sky, Galactic novae thus provide the ideal opportunity to test these predictions in $K$-band light curves. Cryoscope Pathfinder will be sensitive to all Galactic novae in the southern sky, where dust formation can be easily detected, and will reveal a large number of systems that currently elude detection in optical wavelengths \citep{de2021a}. When combined with multi-wavelength data from $\gamma$-ray and radio observatories, these observations can reveal the physics of shocks in these explosions and shed light on the nature of the most luminous shock-powered explosions in the universe such as superluminous supernovae.

{\em Stellar mass black holes}: Searches for stellar mass black holes have been hindered by extinction. Currently, there are only around thirty known stellar mass black holes, even though population synthesis predicts thousands to tens of thousands should be detectable due to photometric modulation with surveys such as {\it TESS} \citep{Masuda2019} and thousands more wide systems due to astrometric measurements from {\it Gaia} \citep{Breivik2017}. In fact, the majority of binary systems containing a black hole are expected to be found at a low scale height in the galactic plane, behind heavy obscuration due to dust, rendering them invisible in the optical \citep[e.g.,][]{Repetto2017}. Thus, to probe this hidden population of objects, one must work at infrared wavelengths which penetrate the dust. A striking example is Cygnus X-3, a high mass X-ray binary containing a Wolf Rayet (WF) star and a candidate black hole \citep{Jones1994} which is invisible in the optical due to extinction, but has a booming 11.91 apparent magnitude in $K$-band. The Pathfinder could be combined with X-ray surveys such as {\it eROSITA} to systematically probe for a complete inventory of such systems by identifying their orbital modulation using their $K$-band light curves.

{\em Stellar populations and Galactic structure}: The purpose of Galactic archaeology is to uncover the formation history of the Milky Way by studying its low-mass, long-lived stars. With cool surface temperatures, these stars mostly emit in the near-infrared. Space-based photometry has been highly successful in determining fundamental stellar properties of red giants through asteroseismology, with significant benefits for Galactic archaeology \citep[e.g.,][]{Casagrande2016}. The potential of detecting and characterizing longer period variables using ground-based $K$-band photometry has been recently demonstrated \citep{Nikzat2022}, opening the possibility to trace intrinsically bright late-type giants across the Galaxy, with a precision exceeding {\it Gaia} beyond 3 kpc \citep{Mathur2016,Huber2017}. As oscillation amplitudes and periods increase with luminosity, stars near the tip of the red giant branch show amplitudes on the order of several parts per thousand with periods greater than 30 days. This allows ground-based telescopes to measure oscillation modes of late-type giants throughout the Galaxy, provided the data covers a long enough time baseline to accurately constrain the period. From space, baseline observations $>30${\days} are only available for a few tens of thousands of giants in the {\it Kepler} field and the {\it TESS} continuous viewing zone \citep{Yu2018,Mackereth2021}. Using simulated synthetic stellar populations from the Galaxia code \citep{Sharma2011}, we estimate about 1 million giants with pulsation periods $> 5${\days} and K $< 12$, thus with enough signal-to-noise to measure oscillations with Cryoscope Pathfinder. A time-domain survey with this facility will allow us to systematically detect and study stellar oscillations in long-period giants across several Galactic components, with major implications for understanding stellar populations in the Milky Way. Stars in common with the {\it TESS} continuous viewing zone will be used to benchmark oscillation detections.

{\em RR Lyrae}: RR Lyrae (RRL) variables are believed to be among the oldest stars in the Galaxy, so are important probes for Galactic archaeology.  They are easy to recognize by their characteristic light curves.  While most RRL stars near the Sun are old ($>10$\,Gyr) and metal-poor ([Fe/H] $< -1$) and are part of the Galactic stellar halo, there is a sub-population of metal-rich RRL stars in the solar neighborhood with abundances [Fe/H] $>-1$ that are clearly part of the Galactic disk. They rotate rapidly around the Galaxy and their velocity dispersion is low.  Recent studies of a sample of a few hundred RRL stars with accurate abundances and kinematics show that these apparently old, metal-rich RRL stars have kinematics similar to relatively young stars of the Galactic thin disk. The origin and evolution of these metal-rich disk RRL stars is not understood. RRL stars of similar [Fe/H] are found in the inner Galaxy, and it is possible that their counterparts in the solar neighborhood have migrated from the inner Galaxy via radial migration. The migration process can generate a kinematically cold population in the outer disk \citep[e.g.][]{VeraCiro2014}.  The radial migration process itself is not well understood. With Cryoscope Pathfinder observations, we will expand our understanding of (i) the metal-rich disk RRL stars near the Sun, and (ii) how radial migration from the inner Galaxy contributes to the stellar populations of the outer disk.

A kinematically cold population of RRL stars in the outer Galactic disk is likely to have a strong concentration to the Galactic plane. So far, there is little data available on RRL star distributions in the low Galactic latitude disk where the interstellar extinction is high. The Cryoscope survey telescope, observing in the $K_{dark}$ band,  would be able to search for these expected low-latitude RRL stars. Such a search can be conducted by observing two Pathfinder fields centered at galactic latitude $b = 0$ and galactic longitude $l$ in the range 280 to 320.  Each field would be observed about 30 times over about a year with irregular cadence. The exposure times for each observation would be long enough to reach a $K_{\rm AB}$\,\,mag of 15, with errors of about 2 percent. The periods of metal-rich RRL stars are mostly in the range 0.35 to 0.5 days. 

\bigskip
\bigskip
\bigskip
\subsection{Gamma Ray Bursts}
Cryoscope's large field-of-view and high cadence in the $K_{dark}$ bandpass will be well-suited for detecting early counterparts to dust-enshrouded and high-redshift gamma ray bursts (GRBs).

GRBs are brief flashes of $\gamma$-rays with an integrated energy spectrum of non-thermal nature, peaking---in a $\nu f_\nu$ representation---around  $\sim$100 keV \citep{Fishman1995}. First serendipitously detected in the late '60s \citep{Klebesadel1973}, it was only 30 years later that arc-minute localization of their X-ray counterparts by the {\it BeppoSAX} satellite led to the long-awaited detection of their multi-wavelength fading remnants, or afterglows \citep{Costa1997,vanParadijs1997,Frail1997}. In particular, optical spectroscopy of the optical counterparts proved their extragalactic origin, with derived observed energy outputs from $10^{48}$\,erg to $10^{54}$\,erg. Their duration distribution is bimodal, with one maximum at $\sim 20$ seconds and another at $\sim 0.2$ seconds \citep{Kouveliotou1993}, leading to a dichotomy between short GRBs, lasting less than 2 seconds with a hard spectrum, and long GRBs, lasting more than 2 seconds and with a soft spectrum. Short GRBs are associated with the mergers of compact objects and long GRBs with the collapse of massive stars.

The detectors flown after \textit{BeppoSAX} on the \textit{HETE-2}, \textit{INTEGRAL} and \textit{Swift} missions continued and perfected the afterglow detection, so that several thousands GRBs have detected afterglows to date, and a large fraction of them have measured redshifts, ranging from 0.008 to $\sim 9.4$.  The detection of GRBs from primordial cosmological epochs suggests their association with the first generation of stars and provides insight into the evolutionary scenario of these massive stellar progenitors.

Observations of long-duration GRBs at low redshifts have established that the vast majority of them are associated with core-collapse stripped-envelope supernovae of high kinetic energy \citep[around $10^{52}$\,erg, e.g. one order of magnitude larger than classical supernovae,][]{Mazzali2021}. Moreover, the concomitant detection of gravitational waves from the binary neutron star coalescence event of 17 August 2017 and of a short-duration GRB (GRB170817A) has definitively proven the association of short-duration GRBs with binary neutron star mergers \citep{MMA}. Recently, however, the dichotomy between long GRBs connected with supernovae and short GRBs associated with kilonovae has become blurred, due to detections of long GRBs without bright supernovae and associations with kilonovae instead \citep{Rastinejad2022,Levan2024},
and short GRBs associated with supernovae \citep{Rossi2022}.

\begin{figure}
  \includegraphics[width=\columnwidth]{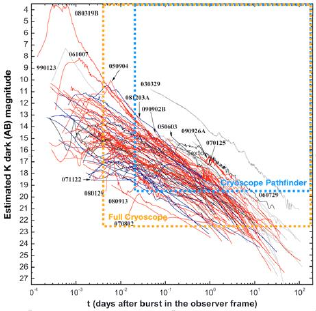}
  \caption{Estimated magnitudes for GRB afterglows in $K_{dark}$ magnitudes derived from the sample of \citet{Kann2011} assuming a spectral slope of $F_\nu$ $\sim 1/\nu$. These light curves are corrected for extinction, which gives us a more realistic view of what is to be expected in $K_{dark}$, where the effect of extinction is less significant. We plot the 5-sigma detection limits for 15 minute exposures with the Pathfinder and the full Cryoscope. Our goal is to aim for a start of the GRB observations within 30 minutes to an hour of the alert with Cryoscope Pathfinder, and within 5 minutes of the alert for the full Cryoscope.
  \label{fig:GRBs}
  }
\end{figure}

A large field-of-view infrared camera with rapid reaction to transients could quickly cover large localization areas on the sky, such as those currently provided by the {\it Fermi}/GBM, {\it Swift}/BAT, {\it SVOM}/ECLAIR and WXT onboard {\it Einstein Probe}, and detect very early counterparts when their brightness and variability amplitude is at a maximum. Observations in $K_{dark}$ are particularly relevant for very high-redshift events at $z > 6$, and for those GRBs that occur deep within dusty regions (see Fig.~\ref{fig:GRBs}). In both cases, the optical light is suppressed, and IR observations are needed to achieve detections. These observations will address critical aspects of GRB science within early observations, such as the interplay of initial emission components (internal versus external shocks; formation of forward and reverse shock; impact and interaction with circumburst medium). At late epochs, the late, non-relativistic transition can be followed, and the unobscured star-formation rate of the host galaxy estimated. Overall, the IR observations provided by Cryoscope will allow us to construct GRB afterglow samples with minimal biases due to extinction and redshift, factors that have strongly affected previous GRB observations.

\subsection{Solar and Interstellar Comets and Asteroids}
    
Small bodies (asteroids and comets) are the remnants of the blocks that accreted to form the planets 4.6\,\,Gy ago. Their importance as witnesses of solar system history has emerged in past decades, but their current description is insufficient to guide theoretical work: current observations cannot discriminate between mutually excluding hypotheses \citep[such as a densely populated or an empty primordial asteroid belt, for instance:][]{2009-Icarus-204-Morbidelli, 2017SciA....3E1138R}. Models may accurately reproduce the dynamical structure of the asteroid belt, but they diverge in their prediction of the distribution of composition. It is therefore crucial to identify which compositions were native to the asteroid belt between Mars and Jupiter and which were implanted later, and what were the implantation mechanisms.

Compositional characterization of a large sample of asteroids can be achieved through broad-band spectro-photometry, complemented with albedo \citep{2013Icar..226..723D, 2014Natur.505..629D}. In these respects, the upcoming NASA \emph{NEO Surveyor} mission \citep{mainzer2023} and the Legacy Survey of Space and Time (LSST) of the Vera C. Rubin Observatory \citep{2009-Book-LSST,2009EMP..105..101J} will play major roles---the former to measure albedo, and the latter to build multiple colors with its six visible-wavelength filters. \emph{NEO Surveyor} will build upon  the discoveries of {\it WISE/NEOWISE} \citep{WISE2010,NEOWISE2011}, which detected $\sim 3000$ NEOs using its 3.4 and 4.6{\um} imaging channels, enabling radiometric modeling of physical properties such as diameter and visible geometric albedo. For optical spectro-photometry with LSST, however, compositional classes of asteroids present a similar spectrum in the visible range and only differ in the infrared \citep{2009Icar..202..160D,2018A&A...609A.113C,2022A&A...665A..26M}. By acquiring photometry in $K$-band at multiple epochs/geometries over several years, Cryoscope Pathfinder will provide for the first time the absolute magnitude and phase function \citep{2010Icar..209..542M} of asteroids in the near-infrared, in a similar fashion as LSST in the visible \citep[the only way to determine the colors of small bodies with LSST, see][]{2021Icar..35414094M}.

The difference of absolute magnitude in the visible and near-infrared will break the spectral degeneracy inherent to only visible observations, providing the compositional distribution of asteroids. Obtaining phase
functions in the near-infrared also opens up new prospects for compositional mapping \citep{2022A&A...665A..26M}: these functions are known to be wavelength-dependent \citep{2012Icar..220...36S} but are seldom acquired with a filter other than Johnson V.

The challenge here resides in the determination of the absolute magnitude, both for Cryoscope and LSST. The photometry of SSOs results from the combination of three varying components, each with a different frequency. The relative motion of the Sun, the target, and the observer implies a change of distances, phase angle (leading to a change of the portion of illuminated surface visible to the observer), and orientation. Moreover, the rotation of the irregular shape introduces an additional intrinsic variability. Observations at phase angles below 3--5$^\circ$ are required to properly solve for the absolute magnitude \citep{2021Icar..35414094M}. The Cryoscope survey would therefore strive to acquire photometry in such configurations (e.g., at a solar elongation of 180$^\circ$).

In addition to the characterization of a sample of asteroids, Cryoscope Pathfinder will be able to discover new asteroids and comets which happen to be in its field-of-view during any scientific observation. Owing to its location in Antarctica, Cryoscope's footprint will be located at high ecliptic latitude, where a deficit of detection of small bodies has been reported \citep{2018AA...610A..21M, 2021A&A...648A..96C}. Newly discovered asteroids of high interest, especially Near-Earth Objects, may be followed up by the Cryoscope or other facilities for improved orbit determination. Among them, objects deemed of particular importance (e.g., potential impactors) could be observed by {\it JWST} for precise measurements of size, orbit, and physical characteristics \citep{2025Natur.638...74B,2023A&A...670A..53M,2025jwst.prop.9239}. However generally, if an asteroid was observed during only a short, one-night arc and was not followed up afterwards, its orbit determination will be poor. Such objects might be lost completely, or until future recovery observations. These one-night arc observations will mostly populate the Isolated Tracklet File of the Minor Planet Center. Such observations might be linked with previous or future observations done by other facilities and will contribute to a pool of objects with well-established orbits. Objects which happen to be in the field-of-view for several consecutive nights might offer better prospects of orbit determination \citep[typically, observations from three different nights are needed for robust orbit determination of a main-belt asteroid, see][and references therein]{2018AJ....156..135H, 2023arXiv230501123L}. Such objects will also be better placed for detection by our synthetic tracking pipeline as more images can be stacked together, increasing the objects' S/N and enabling probes of more distant populations of small solar system bodies fainter than 20\,\,mag \citep{2023MNRAS.521.4568B}.

Interstellar objects (ISOs), which are not gravitationally bound to the solar system, open up an entirely new avenue for placing the solar system in a broader context and understanding planetary system formation more widely. So far, only one interstellar asteroid, ‘Oumuamua \citep{2017Natur.552..378M} and one interstellar comet, 2I/Borisov \citep{2020NatAs...4...53G} have been discovered serendipitously, in 2017 and 2019, respectively. Cryoscope’s vantage point from Antarctica covers the region on the sky containing the solar antiapex. \citet{2022PSJ.....3...71H} suggest that the ISO population is characterized by a clustering of trajectories in the direction of the solar apex and antiapex (the solar apex refers to the direction that the Sun travels with respect to the local standard of rest, with the antiapex directly opposite). Cryoscope’s point of view provides the potential to observe ISOs should they enter the solar system from the direction of the antiapex.

\subsection{Stellar Microlensing}
Cryoscope will occupy a unique niche for gravitational microlensing experiments to detect and characterize populations of dark compact objects within our Galaxy, including primordial black holes and wide-orbit or free-floating planets. As a massive, compact object passes between us and a background star, the effect of gravitational lensing displaces, magnifies, and distorts the observed image \citep{Paczynski1986}. In microlensing, the displacement of the images is on the micro-arcsecond scale, and the image is effectively unresolved. However, the magnification is significant and detectable from a time-series light curve of the lensed source. The amplitude and duration of magnification depend on the geometry and relative motions of the lens-source-observer system. Still, the microlensing timescale is directly proportional to lens mass for a fixed geometry.

With the opportunity to monitor very large stellar fields, Cryoscope is particularly well-suited to microlensing experiments to conduct a gravitational census of compact objects in the Galactic disk and halo.
In particular, with fast-cadence monitoring geared for minute- to day-scale microlensing events, Cryoscope is uniquely suited to probe the mass range $\sim 10^{-12}-10^{-3}$ M$_\odot$, from asteroid- to Jupiter-scale masses. We detail two science cases which will benefit from Cryoscope's survey:

\textit{Wide Orbit and Free-Floating Exoplanets}: Microlensing is a unique probe of `wide orbit' exoplanets and especially free-floating planets (FFPs), which are exoplanets that have been ejected from their birth systems through either planet--star or planet--planet interactions \citep{Mroz2019}. FFP detections are essential for understanding the evolution of planetary systems and determining the ratio of ejected planets to bound ones \citep{Bachelet2022} as a constraint on star- and planet-formation models. KMTNet \citep{Jung2024FFP, YangKMTFFP}, and MOA \citep{Sumi2011FFP} have revealed a substantial population of icy exoplanets and FFPs, with hundreds of detections recorded. Modeling from the 9 years of MOA II data suggests that $19^{+23}_{-13}$ FFPs exist per star \citep{sumi2023}, with masses ranging from terrestrial to Jovian. With its vast field of view and fast integration times, Cryoscope will enable fast-cadence exoplanet microlensing experiments to constrain the number and mass distribution of FFPs in the Milky Way.

\textit{Primordial Black Holes}:
In the same way as for exoplanets, Cyroscope has unique potential for microlensing searches for Primordial Black Holes (PBHs). There is a long-standing hypothesis that a cosmologically significant PBH population may be formed from critically collapsing density perturbations in the pre-inflationary universe \citep{ZeldNov1966, Hawking1971}. Over the last ten years, spurred partly by the possibility that PBHs may explain the unexpectedly large progenitor masses of the LIGO-Virgo mergers \citep{Bird2016}, interest in PBH formation scenarios and possible detection experiments has grown rapidly. Such a population would also be a plausible and natural solution to the dark matter problem, not requiring new particle physics beyond the Standard Model. While there exist some observational constraints on the monochromatic (i.e., uniform mass) PBH density \citep[e.g.,][and references therein]{Carr2024} from microlensing null results, Cryoscope represents an opportunity to determine and/or constrain the Galactic PBH population spanning the asteroid- to Jupiter-mass range, through fast-cadence `sit and stare' monitoring of dark-matter-dominated sight lines towards M31 \citep[e.g.][]{Niikura_2019, Blaineau2022} or the LMC \citep{Mroz2024}.

\textit{Cryoscope's strengths}:
For the microlensing science cases above, the science-limiting factors are: 
(i)\ a vast field of view to allow simultaneous monitoring of the large number of stars; (ii)\ a good and stable atmospheric seeing and transparency to maximize sensitivity to microlensing amplifications and minimize blending and/or contamination; (iii)\ fast integrations and read-out for efficient high-cadence monitoring. Cryoscope delivers on all counts, with the additional benefit of operating in the NIR, where Galactic dust obscuration/attenuation is minimal \citep[e.g.,][]{Shvartzvald2017, Shvartzvald2018, Kondo2023}, and the benefit of very-long night observing for maximally efficient searches on few-hour timescales.

Relative to mega-projects like Rubin, {\it Euclid}, and {\it Roman}, Cryoscope has the opportunity to conduct repeat `sit-and-stare' microlensing experiments over multiple Galactic and extragalactic sight lines. Cryoscope's ability to trace such regions of high stellar
density will increase the probability of detecting multiple microlensing signals, and provide better event statistics to distinguish between the kinematics of baryonic (planetary) and non-baryonic (PBH) microlenses. With the combination of space-competitive seeing and a 4.3$\times$ advantage over DECam in field-of-view, Cryoscope can establish and exploit a unique technical and scientific niche in microlensing experiments, even in the era of Rubin, {\it Euclid}, and {\it Roman}.

\subsection{Core Collapse Supernovae}

Massive stars with $\gtrsim 8~M_\odot$ form an iron core towards the end of their life and die when this iron core exceeds the Chandrasekhar mass and collapses, sending out a shock wave that disrupts the star.
The resulting core-collapse supernova (CCSN) forms a diverse group of transients whose properties are determined by the properties of the progenitor star. 
For instance, the amount of hydrogen left in the envelope determines the spectral type of the CCSN \citep{Filippenko1997}; the star's mass at the time of explosion determines how long the light curve takes to rise \citep{Khatami2019}; and the core mass of the star determines the strengths of different emission lines later in the nebular phase \citep[e.g.,][]{Jerkstrand2014}.
While most of what we know about CCSNe as a class have been derived from optical observations, increasing numbers of IR observations in the past decade revealed several aspects of massive stellar explosions that had remained elusive. 
Cryoscope observations will help us probe the chemical evolution of CCSNe ejecta, reveal late-time power sources of CCSNe, and uncover a population of CCSNe hidden behind dust.

First, CCSNe have complex chemical evolution in their metal-rich ejecta. They initially form simple diatomic molecules with strong chemical bonds, like carbon monoxide and silicon monoxide, then progress to more complicated molecules. 
As the temperature drops, dust grains start to condense, forming either a carbon- or silicate-rich grain population depending on the available species in the ejecta \citep[e.g.,][]{Sarangi2015, Sluder2018, Sarangi2022}.

Modeling such a complex chemical network that produces molecules and dust is crucial for two reasons. 
Molecules and dust dominate the ejecta cooling at late time, and properly accounting for their contribution is crucial in predicting the nebular spectra of CCSNe \citep{Liljegren2020, Liljegren2023}.
Dust from CCSNe can also explain the large observed dust mass at high redshifts, where known local dust producers like AGB stars have yet to evolve \citep[e.g.,][]{Witstok:2023Natur.621..267W, Schneider:2024A&ARv..32....2S}.
The relative contributions of core-collapse supernovae to the cosmic dust budget is an open and hotly debated question. It has been shown that large amounts of dust can form in the ejecta of supernovae \citep[e.g., SN 1987A][]{Wesson:2015MNRAS.446.2089W, Shahbandeh2023} but supernovae may also destroy large quantities of dust when they explode \citep{Priestley:2021MNRAS.500.2543P}. In particular, the reverse shock may destroy dust that is newly formed in the ejecta of the supernova \citep{Micelotta:2016A&A...590A..65M}.
This question of dust destruction is also essential to understanding the contribution of supernova progenitors (e.g., red supergiants---RSGs) as it is known they can form modest quantities of dust but it is highly uncertain how much will actually survive the supernova explosion. In addition, understanding how much pre-existing CSM dust survives the supernova forward shock also informs our understanding of dust destruction by supernovae: both the destruction of ISM dust by the forward shock and newly formed dust in the ejecta by the reverse shock.
It is also not known how easily dust reforms in the shocked material after it is destroyed. 
Cryoscope time series NIR observations of dust emission, both pre-existing and newly formed, will allow us to study this process of formation, destruction, and possible reformation and provide important insights into the contribution of core-collapse supernovae to the cosmic dust budget.

While {\it JWST} observations are extremely powerful in tracking molecule and dust formation in CCSNe, the number of observations is still limited to a few nearby and notable objects \citep[e.g.,][]{Hosseinzadeh2023, Shahbandeh2023}. 
Cryoscope is unique as the only planned ground-based, high-cadence survey that covers wavelengths beyond 2{\um} (unlike the DREAMS and WINTER surveys powered by InGaAs detectors).
The $K_{dark}$ band, unlike the more typical $K_s$, covers the bandhead of CO's first overtone band at around 2.3{\um}. 
$K_{dark}$ observations will also be able to track the rising thermal continuum from newly formed dust grains in the ejecta of CCSNe. 
Unlike in previous targeted IR observations, Cryoscope will observe a statistically significant sample of CCSNe and measure the evolution of CO and dust formation to test chemical evolution models of CCSNe.

Second, IR observations are crucial to monitor late-time evolution of CCSNe bolometric luminosity, which indicate the powering mechanism. 
As the ejecta cool several months to years post-explosion, the peak of the SED progressively shifts into the IR, and observations only in the optical quickly miss the majority of the flux.
This is largely due to the aforementioned dust formation, which absorbs optical photons and re-emits them in the IR.
An understanding of the bolometric luminosity at these epochs is crucial to unveil additional powering mechanisms in CCSNe as the primary power from radioactivity decays away.

The most common additional powering mechanism observed in CCSNe is interaction between the SN shock and the circumstellar medium (CSM) ejected by the progenitor star prior to the explosion. 
While some CCSNe interact with dense and nearby CSM, producing luminous SNe with narrow emission lines from the ionized, slowly moving CSM (SNe IIn; \citealp{Schlegel1990}), late-time observations have shown that other typical CCSNe show signs of faint CSM interactions years post-explosion. 
These late-time CSM interactions probe mass loss in massive stars at earlier time prior to explosion \citep{Myers:2024ApJ...976..230M} and can provide important insights into massive stellar physics in the hundreds of years prior to supernovae.

In Type II-P SNe, late-time CSM interaction can measure the steady mass loss rate in RSGs, a hotly debated number in the massive star community \citep[e.g.,][]{Beasor2020}.
Several SNe II-P have been observed with IR excess several years post-explosion, attributed to the SN shock interacting with the progenitor steady winds. 
Examples include SNe 1980K \citep{Zsiros2024}, 2004dj \citep{Meikle2011, Szalai2011}, 2004et \citep{Kotak2009, Maguire2010, Fabbri2011, Shahbandeh2023}, and 2017eaw \citep{Weil2020, Shahbandeh2023}. 
Cryoscope will be the first systematic survey that can determine the rate at which SNe II-P show late-time interactions.

Another class of objects for which late-time interaction rate is crucial to measure is stripped-envelope (SE) SNe. 
These are CCSNe from a progenitor star that lost most or all of its hydrogen-rich envelope (Type IIb and Ib, respectively) or even most of the helium-rich envelope (Ic). 
Single-star wind-driven mass loss cannot explain either the rate or the low progenitor mass observed in most SESNe, and binary-driven mass loss likely produces most of SESNe progenitors \citep[e.g.,][for a review]{Smith2014}.
To test this hypothesis, direct observations of CSM around SESNe are crucial to measure the density profile and mass-loss rate to determine its origin. 
Thus far, very few SESNe have been observed interacting with their lost envelope, most notably SN 2014C, which starts to interact with a detached CSM a few months post-explosion and is still ongoing \citep{Milisavljevic2015, Margutti2017, Tinyanont2019}. 
The rarity of interacting SESNe suggests that the time delay between the conclusion of envelope stripping and core collapse is likely substantial.

While CSM interactions have many emission signatures, dust IR emission is the most sensitive; weak interactions that do not produce detectable X-ray or radio emissions or narrow optical lines still heat enough dust to temperatures at which they emit in the IR.
Therefore, Cryoscope will be a powerful tool to detect or constrain late-time interactions in SESNe. 
Its systematic nature will provide the first IR measurement of the rate of interactions at different times post-explosion. 
This number relates to the time delay between envelope stripping and core collapse, which will precisely put the mass loss event on the stellar evolutionary timescale.

\subsection{High-energy neutrino counterparts}

A flux of high-energy neutrinos was first discovered by IceCube in 2013 \citep{ic_astro_2013}, and while we now know that this flux has both extragalactic and galactic components \citep{ic_gp_23}, the origin of this flux remains mostly unknown. Public neutrino `alerts' are now routinely released by neutrino telescopes in low latency \citep{ic_realtime},
leading to the association of individual neutrinos with flaring blazars \citep{txs_mm} and Tidal Disruption Events (TDEs) \citep{stein21,reusch2022,vanVelzen:2021zsm}.

Cryoscope will search for infrared counterparts of high-energy neutrino alerts in the southern hemisphere. Cryoscope fills a crucial gap in follow-up capability because there are no other instruments capable of sensitive searches for galactic transient or variable counterparts despite the recent discovery of neutrinos from the Galactic plane \citep{ic_gp_23}. The impact of galactic extinction means that such searches in the UV or optical are extremely challenging. As an infrared telescope, Cryoscope will be substantially more sensitive than optical telescopes for discovering neutrino counterparts that suffer from heavy galactic extinction, as well as those which are intrinsically red (e.g., GRB afterglows), or are dust-obscured (e.g., TDEs or CCSNe). In this sense, Cryoscope is truly unique and will enable us for the first time to comprehensively search for proposed time-varying galactic neutrino sources such as novae \citep{metzger_16,fang_20,bednarek_22}.

While only a handful of IceCube neutrino alerts will be accessible to Cryoscope, most alerts from northern large-volume neutrino telescopes will be accessible, such as the KM3NeT telescope\footnote{https://www.km3net.org/}~\citep{KM3Net2016} and the Baikal GVD telescope~\citep{Avrorin2022}. In addition, other facilities are also planned in the future such as P-ONE~\citep{PONE2020} in Canada or TRIDENT~\citep{Ye2023} in China. As water-based neutrino detectors, all of these northern telescopes will have superior angular resolution to IceCube (due to the increased light scattering in ice-based detectors). Moreover, since neutrino telescopes are most sensitive to events approaching the detector horizontally or from overhead, these northern neutrino telescopes will produce far more alerts from the Galactic plane and southern hemisphere.

Cooperation between Cryoscope and neutrino facilities holds the promise of complementary operation. For instance, KM3NeT intends to notify the astronomical community upon detecting noteworthy events, including those resulting from electron and tau neutrino interactions. 
In this context, having a wide field-of-view is beneficial, as these events are localized in the sky with a median angular error of over 1 degree. We anticipate receiving approximately two neutrino alerts per month from KM3NeT. 
We propose to follow up two neutrinos each month, beginning with an initial observation in the first hour after neutrino detection, and two subsequent visits over the next month. This approach benefits from the extensive experience of our team members, who have employed this strategy for several years \citep[refer to][for details]{Dornic2009}. This strategy provides sensitivity to a broad range of potential neutrino source classes, which evolve on timescales from hours (e.g., GRB afterglows) to months (e.g., TDEs). 

Neutrino follow-up will benefit from the very low-latency observations which Cryoscope can perform during Antarctic winter.
Moreover, it will benefit from substantial synergy with planned optical follow-up of southern neutrinos with the Vera Rubin Observatory \citep{Andreoni2024}.

Cryoscope also provides a new opportunity to explore extragalactic TDEs as potential sources of neutrinos, with three TDEs exhibiting infrared dust echoes coincident with neutrino detections \citep{vanVelzen:2021zsm}. These IR echoes were only detected by the former {\it NEOWISE} satellite \citep{Mainzer2014}, which ceased survey operations in July 2024, leaving a critical gap which Cryoscope will soon fill. 
To systematically investigate this, we will identify any previously-reported TDEs in the error region of each neutrino alert and check for the ongoing dust echoes detected during follow-up. If any infrared counterpart is found during the first visit, we will revisit these TDEs over the following weeks. 
This approach is tailored to optimize the efficiency of investigating TDEs while simultaneously minimizing the telescope observation time required.
Thus, even with the Pathfinder, we can investigate what fraction of high-energy neutrinos is associated with supermassive black hole activity and what fraction shows infrared excesses.

Besides high-energy neutrino events, various neutrino telescopes such as Super-Kamiokande, JUNO and KM3NeT are actively surveilling the sky for low-energy neutrinos originating from the next galactic or nearby core-collapse supernova. These telescopes are all part of the SNEWS project~\citep{Habig:2005ooj}. If at least two neutrino detectors issue an alert to SNEWS within a coincidence time window of less than 10 seconds, a public alert with an exceptionally low false alarm rate is promptly disseminated. In the recent upgrade, SNEWS 2.0~\citep{SNEWS:2020tbu}, real-time information will be provided concerning localization, potentially obtained either from individual neutrino detectors or through triangulation using signals from multiple neutrino telescopes, along with a distance estimation and certain supernova characteristics. The validation of a neutrino signal's origin from a core-collapse supernova relies on near-infrared observations, necessitated by the significant line-of-sight extinction toward the Galactic center. Consequently, Cryoscope's role will be indispensable in this follow-up process.

\subsection{Supermassive Black Holes}
Sitting at the heart of nearly every galaxy is a supermassive black hole (SMBH) that can be millions to billions times the mass of the Sun. These reveal themselves across the electromagnetic spectrum through interactions with stars and gas in their environment: when a star passes within the sphere of disruption of the SMBH, tidal forces will tear the star apart with a flash of radiation known as a tidal distruption event \citep[TDE;][]{rees:1988Natur.333..523R}, while gas funneling in can form an optically thick, geometrically thin accretion disk around the SMBH creating an active galactic nuclei (AGN). SMBHs can be energetically dominant in their host galaxies and even dictate how the galaxy evolves. 
Variability is a key characteristic of SMBHs and has been extensively studied in optical and X-ray wavelengths, but IR studies have been largely limited to the six-month cadence decadal data set that {\it WISE} provided at mid-infrared wavelengths. IR variability probes the outer regions of the accretion disk and the dusty torus around the SMBH and can feature significant contribution from the host galaxies, providing important insights into these environments.

While the vast majority of TDEs to date have been discovered based on their X-ray \citep{Saxton:2020SSRv..216...85S} or UV/optical emission \citep{vanVelzen:2020SSRv..216..124V}, IR wavelengths offer critical diagnostics that cannot be gleaned from the current sample. In particular, Cryoscope observations will be exquisitely sensitive to the presence of dust echoes from the initial X-ray/UV emission \citep{vanVelzen:2021SSRv..217...63V}---in fact the Cryoscope $K_{dark}$ bandpass directly samples the peak of thermal emission expected for the dust heating ($T \sim 1800$\,\,K). Based only on the observed sample of UV/optical TDE discoveries, with a 10,000{\sqdeg} survey starting when the direct drive becomes available in year 2, we expect Cryoscope Pathfinder to discover $\sim$ a dozen new IR-selected TDEs each year. By comparing the properties of the IR dust emission with the UV/optical emission, we can directly measure both the dust location (based on the lag of the IR signal) as well as the dust-covering fraction \citep[based on the ratio of IR to UV/optical luminosity;][]{Jiang:2021ApJ...911...31J}. Such measurements are exceedingly difficult to obtain by any other means, providing a completely unique picture of the geometry of the SMBH environment in distant, quiescent galaxies.

Several hundred AGN have been identified in the past decade that exhibit both strong photometric and spectroscopic optical variability known as changing-state AGN \citep[CS-AGN;][]{Ricci:2023NatAs...7.1282R}. The physical mechanism is not clear but seems to be related to an accelerating outflow or varying accretion rates onto the SMBH, which may in turn be driven by physical changes in the accretion disk or its magnetic field support. IR variability is seen to echo the optical variability with large amplitude variations and a time lag expected for dust reprocessing. Cryoscope enables novel studies of CS-AGN with much higher cadences. Cryoscope will be able to probe the interface region between the accretion disk and the torus; identify higher redshift CS-AGN where rest-frame optical continuum changes have been redshifted into the observed near-IR frame; and detect near-IR CS-AGN where the optical variability may be obscured by dust. There is also the possibility of capturing CS-AGN where a triggering event occurs on the outskirts of the accretion disk and a thermal front then propagates radially inwards, temporarily altering the disk geometry.  

Supermassive black hole binaries (SMBHBs) are an expected consequence of hierarchical models of galaxy formation and an important source of nanohertz frequency gravitational waves (detectable by {\it LISA}). At close separations, they are not physically resolvable, but induced effects from the kinematics of the rotating pair may be imprinted on electromagnetic (EM) radiation from the SMBHB. The two main physical mechanisms for an EM signature are: (i) relativistic Doppler boosting of emission from a mini-disk associated with one of the binary; and (ii) periodic modulation of the accretion rate on the SMBHB. A significant distinction between the two is that variability due to accretion rate fluctuations should be isotropic, whereas relativistic Doppler modulation produces anisotropic variability resembling a rotating forward-beamed lighthouse surrounded by fog. This can have observational consequences for IR variability depending on the geometry of the AGN system, e.g., the ratio of dust light crossing time to the source variability period, torus inclination, and opening angle. In particular, there should be a population of nearly face-on binaries with an inclined dust torus and modulated IR variability but not observable optical/UV variability that would have been missed in optical searches, but which Cryoscope has the sensitivity and field coverage to easily find.

Finally, the most powerful technique to probe AGN is reverberation mapping (RM), where measured light travel time delays map out spatial scales that are often well beyond imaging capabilities of current telescopes. The RM technique was pioneered in the optical where the location and motion of gas flows that constitute the broad-line region are constrained by observing the time-delayed response of broad emission lines to the variable continuum flux from the accretion disk. Near-IR RM tracks emission produced in the dusty torus: the dust is irradiated by X-ray and UV/optical continuum emission from the inner regions and is reprocessed as thermal emission in the $1300-2000$ K range. The dust only survives at temperatures below the sublimation temperature, and therefore this temperature defines the inner radius of the dusty torus. As the torus is much larger and further out than the inner accretion disk, the IR variability is longer and weaker, and the corresponding reverberation lags are longer. Cryoscope again provides the capability for extensive near-IR monitoring of the AGN population, which coupled with optical light curves from Rubin will create an unprecedented sample of optical-to-IR lags over a range of black hole masses, accretion rates, and orientations. 

\section{Summary and Path Forward}\label{sec:summary}
Opening a new window into our dynamic infrared sky presents an opportunity to discover fast transients such as kilonovae, overcome the time-intensive requirements on exoplanet detection and exo-atmospheric characterization, and unravel facets of time-domain, stellar, and solar system science that need long-wavelength coverage. New surveyors are emerging in the coming years that will uncover the changing universe in the ultraviolet, optical, high-energy and radio. However, there is no analogous project in the infrared that can simultaneously provide deep, wide, and fast coverage of the continuously evolving night sky. There is no infrared surveyor that is capable of tiling $\approx$ 20,000{\sqdeg}, at hourly cadences, below 20th AB mag, and at wavelengths beyond 2{\um}. Such a sensitive, high-cadence surveyor in the infrared would unveil a new phase space for exploration and, thus, new possibilities for discoveries. 

To this end, we are developing a new telescope that will provide the coverage needed for discoveries in dynamic infrared astronomy. Cryoscope uses a novel optomechanical design to keep its optical path at cryogenic temperatures, eliminating thermal background from the ground within the $K_{dark}$ bandpass (2.25-2.5{\um}). This allows the instantaneous field-of-view to be increased $80\times$ that of other previous deep surveyors in the $K$-band (e.g., VISTA), at no cost to its sensitivity. To image its large focal plane, Cryoscope will employ a new infrared camera with more than 600 Mpix, the highest number of pixels ever made for infrared astronomy. To achieve the depth demanded for the greatest variety of science cases, Cryoscope will be uniquely situated at Dome C, Antarctica, relying upon infrared skies that are at least 30 times darker than the best astronomical sites at temperate latitudes (e.g., Mauna Kea, Cerro Pachón). Dome C is already home to the productive ASTEP+ project, which has complemented {\it TESS} to propel exoplanet science in the optical bands. We now seek Dome C's infrared sky. Reduced sky brightness in $K_{dark}$ over the elevated Antarctic plateau, coupled with clear weather, unparalleled seeing, and continuously dark winter nights, makes the location of Dome C critical for any ground-based, high-cadence, deep and wide infrared surveyor. 

Cryoscope represents an ambitious leap to answer fundamental questions in the infrared, closed off for decades due to technical obstacles. Ground-based infrared telescopes have not yet mitigated the threat of high thermal background noise in the $K$-band when expanding their fields-of-view beyond 1{\sqdeg}, and have not been able to overcome the bright sky backgrounds from other ground-based sites. Moreover, the high cost of infrared detector technology has precluded the push to tile the focal plane delivered by a truly wide field-of-view infrared surveyor.

We have built a fifth-scale prototype of the full-scale telescope, the Cryoscope Pathfinder, to demonstrate that these obstacles are surmountable and these science cases are within reach. Due to be deployed in December 2026, the Pathfinder will for the first time measure the $K_{dark}$ sky brightness at Dome C during the winter observing season and retire the technological risk of operating a fully cryogenic telescope in the Antarctic. Cryoscope Pathfinder is currently being tested in the lab and has been shown to deliver diffraction-limited imaging across its field-of-view. Upcoming tests will soon validate the thermal management strategy to minimize the telescope thermal emission well below the dark sky levels conservatively predicted at Dome C. On-sky tests will then demonstrate focus stability and verify the image quality revealed in the lab.  

At Dome C, successful commissioning of the Pathfinder will entail ensuring survival during operation, maintaining thermal control over the detector and optics core, and preventing condensation on the window corrector lens at the entrance aperture. We will demonstrate our ability to manage the volume of data produced by a wide-field infrared survey and develop survey strategies in preparation for the full-scale system. To take advantage of the $<0.3$-arcsec seeing above a 20{\m} inversion layer from ground-level at Dome C, Cryoscope will be installed on a vibration-isolating mount in a dome mounted on a 25{\m} tower. In future work, we will detail the design of such an observatory that will house the full-scale Cryoscope and meet the logistical challenges underpinning its Antarctic expedition. 

In summary, the Cryoscope project will unveil the dynamic infrared sky with its unmatched volumetric survey speed. Driven by new optical-thermal designs for wide-field telescopes, advances in infrared detector technology, and the darkest infrared skies on the planet, Cryoscope will serve as a discovery engine for explosive transients in multi-messenger astronomy, stellar variables, and exoplanets around cool stars.  

\bigskip
\bigskip
\noindent {\bf Acknowledgments} \\
The Cryoscope Pathfinder project is supported by Schmidt Sciences (12540478). We also acknowledge support from the National Science Foundation (ATI 2010041, RAPID 2449325) and the Mount Cuba Astronomical Foundation (12540465). We acknowledge support for detector development from NSF (AST 1509716) and NASA (NNX13AH70G). We thank IPEV for supporting our on-site operations in Concordia, Antarctica. 

\software{\texttt{astropy} \citep{astropy:2013, astropy:2018, astropy:2022}, \texttt{matplotlib} \citep{Hunter:2007},
\texttt{numpy} \citep{harris2020array},
\texttt{pandas} \citep{2022zndo...3509134T},
\texttt{scipy} \citep{2020SciPy-NMeth}, \texttt{sncosmo} \citep{2022zndo....592747B},
\texttt{synphot} \citep{synphot2018}}

\appendix
\section{Volumetric survey speed across the UVOIR}\label{appendix}
\restartappendixnumbering 
Volumetric survey speed is a figure of merit detailing the rate at which a surveyor tiles the universe to its limiting depth, folding in a survey's instantaneous field-of-view with the exposure and overhead times needed to detect sources at a given absolute magnitude. From \citet{Bellm2016}, volumetric survey speed is given by 
\begin{equation}
    \dot V = \frac{\Omega_{FOV}}{4\pi}\frac{V_c [z_{lim}(M, m)]}{t_{exp}+t_{OH}}
\end{equation}
where $\Omega_{FOV}$ is the instantaneous field-of-view in steradians, $V_c$ is the co-moving volume of the universe in Mpc$^3$ accessible at a limiting redshift $z_{lim}$, which is further parametrized by a given absolute magnitude $M$ and limiting apparent magnitude $m$. In the calculations that follow we presume $M=-19$, typical for peaking type Ia supernovae, and limiting apparent magnitudes at the $5\sigma$ detection level for a point source. In the calculation of $V_c$, we use standard flat $\Lambda$CDM cosmology with $H_0 = 70.4${\km}\,s$^{-1}$\,Mpc$^{-1}$, $T_{CMB} = 2.725$\,K and $\Omega_0 = 0.272$. To determine the comoving volume, we numerically find $z_{lim}$ for the given depth, applying a K-correction of $K = -2.5\log(1 / (1 + z))$, and use {\tt astropy.cosmology} to calculate the comoving volume at the redshift. 
The visit time is the sum of exposure and overhead times ($t_{exp}+t_{OH}$). Nominal visit times include dither sequences, if relevant.

\startlongtable
\centerwidetable
\begin{deluxetable}{p{2cm}cccccc}
\tablecolumns{7}
\tablecaption{Selection of UVOIR nominal survey properties \label{tab:landscape}}
\tablehead{
\colhead{Survey} & 
\colhead{Year} & 
\colhead{FoV (deg$^2$)} & 
\colhead{Visit time (s)} & 
\colhead{Filter} & 
\colhead{$5\sigma$ AB depth/visit} & 
\colhead{Survey speed (Mpc$^3$/s) } \\
& & & ($t_{exp}+t_{OH}$) & & & $\dot V (M_{AB} = -19)$
}
\startdata
    UVEX & Planned 2030 & 12.25 & 900  &  NUV-FUV & 24.5 & $1.84\times10^4$\\ [0.5em]
    \hline \\ [0.5em]
    ULTRASAT & Planned 2027 & 204 & 900 & NUV &  22.5 & $5.08\times10^4$\\ [0.5em]
    \hline \\ [0.5em]
    GALEX All-sky Imaging Survey & 2003-2007 & 1.1 &  100  &  NUV-FUV & 20.5 & $2.92 \times10^2$\\ [0.5em]
    \hline \\ [0.5em]
    Rubin LSST & First light 2025 & 9.6 & 2$\times$15+9 &  $u$ & 23.9 & $2.01\times10^5$\\ [0.5em] 
    & & & & $g$ & 25.0 & $4.92\times10^5$\\ [0.5em] 
    & & & & $r$ & 24.7 & $3.90\times10^5$\\ [0.5em] 
    & & & & $i$ & 24.0 & $2.19\times10^5$\\ [0.5em]
    & & & & $z$ & 23.3 & $1.18\times10^5$\\ [0.5em]
    & & & & $y$ & 22.1 & $3.70\times10^4$\\ [0.5em] 
    \hline \\ [0.5em]
    ZTF & 2018-present & 46.7 & 30+10 &  $g$ & 20.8 & $4.36\times10^4$\\ [0.5em]
    & & & & $r$ & 20.6 & $3.48\times10^4$\\ [0.5em]
    \hline \\ [0.5em]
    Roman HLTDS, Deep Tier & First light 2027 & 0.281 & 300+70 & F106 & 27.5 & $7.68\times10^3$\\ [0.5em]
    & & & & F129 & 27.4 & $7.26\times10^3$\\ [0.5em]
    & & & & F158 & 27.3 & $6.86\times10^3$\\ [0.5em]
    & & & 900+70 & F184 & 27.5 & $2.93\times10^3$\\ [0.5em]
    \hline \\ [0.5em]
    Roman GBTDS & First light 2027 & 0.281 & 120$^*$ & F087 & 24.6 & $3.41\times10^3$\\ [0.5em]
    &&&& F146 & 25.5 & $6.82\times10^3$\\ [0.5em]
    &&&& F213 & 23.3 & $1.10\times10^3$\\ [0.5em]
    \hline \\ [0.5em]
    Euclid NISP, ROS & 2023-present & 0.55 & 4396$^*$ & $Y_E$ & 24.3 & $1.43\times10^2$\\ [0.5em] 
    & & & & $J_E$ & 24.5 & $1.69\times10^2$\\ [0.5em]
    & & & & $H_E$ & 24.4 & $1.56\times10^2$\\ [0.5em]
    \hline \\ [0.5em]
    Palomar-Gattini IR & 2019-present & 24.7 & 8$\times$(8.1+4.9) & $J$ & 15.7 & 14.6\\ [0.5em]
    \hline \\ [0.5em] 
    WINTER & 2023-present & 1.2 & 8$\times$(120+5) &  $J$ & 19 & 5.3\\ [0.5em]
    \hline \\ [0.5em]
    VISTA Hemisphere Survey & 2009-2013 & 0.6 & 60 & $J$ & 21.1 & $5.21\times10^2$\\ [0.5em]
    & & & & $K_s$ & 19.9 & $1.32\times10^2$\\ [0.5em]
    \hline \\ [0.5em]
    SPHEREx All-sky Survey & First light 2025 & 12.25$^*$ & 112 & Band 1 (0.75--1.11{\um}) & 19.517$^*$ & $9.15 \times 10^2$\\ [0.5em]
    &&&& Band 2 (1.11--1.64{\um}) & 19.522 & $9.20 \times 10^2$\\ [0.5em]
    &&&& Band 3 (1.64--2.42{\um}) & 19.6 & $9.73 \times 10^2$\\ [0.5em]
    &&&& Band 4 (2.42--3.82{\um}) & 19.8 & $1.23 \times 10^3$\\ [0.5em]
    &&&& Band 5 (3.82--4.42{\um}) & 18.7 & $3.30 \times 10^2$\\ [0.5em]
    &&&& Band 6 (4.42--5{\um}) & 18.1 & $1.50 \times 10^2$\\ [0.5em]
    \hline \\ [0.5em]
    Cryoscope Pathfinder & Planned 2026 & 16.2 & 8$\times$(10+5) & $K_{dark}$ & 18.0 & $1.80\times10^2$\\ [0.5em]
    \hline \\ [0.5em]
    Cryoscope & Proposed 2030 & 49.6 & 8$\times$(10+5) & $K_{dark}$ & 21.9 & $5.24\times10^4$\\ [0.5em]
    \hline \\ [0.5em]
    WISE All-sky survey & 2010 & 0.61 & 8$\times$(7.7+3.3) & $W1$ & 19.2 & 37.9\\ [0.5em] 
    & & & & $W2$ & 18.8 & 24.8\\ [0.5em] 
    & & & 8$\times$(8.8+2.2) & $W3$ & 16.4 & 1.1\\ [0.5em]
    & & & & $W4$ & 14.5 & 0.08\\ [0.5em]
    \hline \\ [0.5em]
    NEO Surveyor & Planned 2027 & 11.89 & 6$\times$30 & NC1 (4--5.2{\um}) & 19.4$^*$ & $4.61\times10^2$\\ [0.5em]
    & & & & NC2 (6--10{\um}) & 18.8$^*$ & $2.29\times10^2$\\ [0.5em] 
\enddata
\tablerefs{
{\bf UVEX}: Table 1, \citet{Kulkarni2021}.
{\bf ULTRASAT}: Table 1, \citet{ultrasat2024}.
{\bf GALEX AIS}: Table 1, \citet{galex2005}.
{\bf Rubin}: Table 2, \citet{Bianco2022}, using science-required limited depths.
{\bf ZTF}: \citet{Bellm2019}.
{\bf Roman}: Two reference surveys are considered, the high latitude time domain survey (HLTDS) and Galactic bulge time domain survey (GBTDS). A selection of the reddest filters are chosen from HLTDS in \citet{Rose2021}. $10\sigma$ depths presented in Table 2 are adjusted to $5\sigma$ depths for the same exposure time, assuming background-limited observations: $m_{5\sigma} \approx m_{10\sigma}+2.5\log2$. $^*$For proposed concepts of GBTDS, the nominal survey cycles through $\sim$ 2{\sqdeg} total area every 15\,min. We assume $15/(2/0.281)\sim 2$\,min visits including overheads with nominal 73\,s exposures. The primary filter to be used will be the broadband F164, with supplemental coverage via F087 and F213 at different cadences. Filter combinations have not been finalized (\url{https://asd.gsfc.nasa.gov/roman/comm_forum/forum_17/Core_Community_Survey_Reports-rev03-compressed.pdf}).  We approximate the limiting depth reached in these stares in comparison to depths accomplished in 57\,s background-limited observations: $m_{73\rm{s}} \approx m_{57\rm{s}} + 1.25\log(73/57)$ (for depths in 57\,s single exposures, see \url{https://roman.gsfc.nasa.gov/science/WFI_technical.html}).
{\bf Euclid}: $^*$See \citet{Euclid2022}, Section 4 for the breakdown between overhead and exposure time for a given visit in the reference observing survey (ROS) and Table 7 for $5\sigma$ depths. Only considering NISP and the depths reached therein, yet across the entire NISP FoV instead of the shared NISP+VIS FoV (0.55 instead of 0.53{\sqdeg}). 
{\bf Palomar Gattini-IR}: \citet{PGIR}. 
{\bf WINTER}: Internal communication at Caltech, nightly survey operations currently in $J$-band. 
{\bf VISTA}: \citet{vhs2013,VISTA}, Active field of view is 0.6{\sqdeg}.
{\bf SPHEREx}: $^*$Given {\it SPHEREx}'s linear-variable filters (LVF), we employ the median depth and survey speed across the wavelength coverage in each of the six detector bands. Each of the six detector bands, with $3.5^\circ \times 3.5^\circ$ field-of-view, is divided into 17 spectral channels \citep{Hui:2024SPIE13092E..3NH}. All-sky point-source sensitivities in each spectral channel are obtained from {\it SPHEREx} public products (\url{https://github.com/SPHEREx/Public-products}). 
{\bf Cryoscope+Pathfinder}: This work.
{\bf WISE}: Sensitivities in \citet{WISE2010} correspond to co-added stacks of 8 exposures. We define the 8 exposures to encompass a visit.
{\bf NEO Surveyor}: $^*$Assuming monochromatic flux densities of 65 and 110\,\,$\mu$Jy \citep{mainzer2023}.
}
\end{deluxetable}

\section{First principles argument for a cryogenic optical path}\label{appendix:cryobasis}
The differential energy being transferred to the central pixel is 
\begin{equation}
    dE = I_\lambda dA d\Omega dt d\lambda
\end{equation}
where $I_\lambda$ is the specific intensity, $A$ is the area element radiation is being transferred to, $d\Omega$ is the solid angle subtended to the pixel, $t$ is time, and $\lambda$ wavelength. 

The differential power delivered to the central pixel is given by 
\begin{equation}
    dP = \frac{dE}{dt} = (I_\lambda d\lambda) dA d\Omega
\end{equation}

Assuming uniform blackbody emission and integrating over the bandwidth of the filter, area, and solid angle yields

\begin{equation}
    P = B(T) A \Omega
\end{equation}

Thermal self-emission is thermal emission from the telescope itself. The walls are assumed to dominate, being the closest approximate to a pure blackbody and given the low emissivity of other components in the system such as the fused silica lenses. 

The power delivered to the central detector pixel from the wall's thermal emission is given by 
\begin{equation}
    P_{wall} = B(T_{wall}) n_p^2 \Omega_{wall}
\end{equation}
where $B(T_{wall})$ is the Planck function parametrized by the temperature of the walls, $n_p$ is pixel size, and $\Omega_{wall}$ is the solid angle subtended by the walls at the central pixel.

We approximate the optics core as a cylinder. At a given point on the wall, 
\begin{equation}
    r^2 = z^2 + \rho^2
\end{equation}
where $r$ is the distance from the central detector pixel to the wall, $z$ is the axial height above the origin, defined as the detector plane, and $\rho$ is the distance from the optical axis to the wall. $\rho = D/2$ where $D$ is the aperture diameter.

The differential area element of a patch of the wall in cylindrical coordinates, $dA_p$, is given by: 
\begin{equation}
    dA_p = \rho dz d\phi
\end{equation}

Therefore, the differential solid angle element is given by 
\begin{equation}
    d\Omega = \cos\theta \frac{dA_p}{r^2} = \left(\frac{\rho}{\sqrt{z^2+\rho^2}}\right) \left(\frac{\rho dz d\phi}{z^2 + \rho^2}\right) = \frac{\rho^2 dz d\phi}{(z^2+\rho^2)^{3/2}}
\end{equation}

where the factor of $\cos\theta$ accounts for the angle of incidence between the normal $\hat{n}$ of the area element on the wall, defined to point radially inwards towards the optical axis, and the position vector towards the detector, $\hat{r}$.

We assume azimuthal symmetry and integrate over $z$ from the minimum height above the origin $z_{min} = \rho \tan \theta_{min}$, in which a ray from the wall is able to clear the detector housing and where $\theta_{min}$ is determined from the geometry of the detector housing, to the focal length of the telescope, $f$.  
\begin{equation}
\Omega_{wall} = 2\pi \rho^2 \int_{z_{min}}^{f} \frac{dz}{(z^2 + \rho^2)^{3/2}} 
\end{equation}

For the Pathfinder, an aperture diameter of $26$\,cm, and a focal ratio of $F=2$, $\Omega_{wall} \sim 1.5$\,steradians.

To compare with thermal self-emission from the telescope walls, we roughly assume blackbody emission from both sky and telescope. The power from the sky: 
\begin{equation}
    P_{sky} = B(T_{sky}) \pi \left(\frac{D}{2}\right)^2 \left(\frac{n_p}{f}\right)^2
\end{equation}
where $B(T_{sky})$ is the Planck function given by the temperature of the sky, $f$ is focal length, and $D$ the diameter of the primary mirror. The collecting area is now the area of the primary mirror rather than just the area of the central pixel element. The solid angle subtended by the sky to the detector is the plate scale squared, $(n_p/f)^2$.

Comparing thermal emission between the two dominant contributions: 
\begin{equation}
    \frac{P_{wall}}{P_{sky}} = \frac{B(T_{wall})}{B(T_{sky})} * \Omega_{wall} * \left( \frac{n_p}{\frac{n_p}{f}} \right)^2 * \frac{1}{\pi \left(\frac{D}{2}\right)^2} = \frac{B(T_{wall})}{B(T_{sky})} * \Omega_{wall} * \left(\frac{f}{D} \right)^2 * \frac{4}{\pi}
\end{equation}

Assuming the temperatures of the wall and sky are roughly equivalent: 

\begin{equation}
    \frac{P_{wall}}{P_{sky}} = \frac{16 \Omega_{wall}}{\pi}
\end{equation}

The power delivered to the detector from the Pathfinder's thermal self-emission is then roughly $10\times$ the power delivered from the sky's thermal emission. In order for sky brightness to be the dominant background term, the telescope always needs to be colder than the sky.

\begin{figure}
\centering
\includegraphics{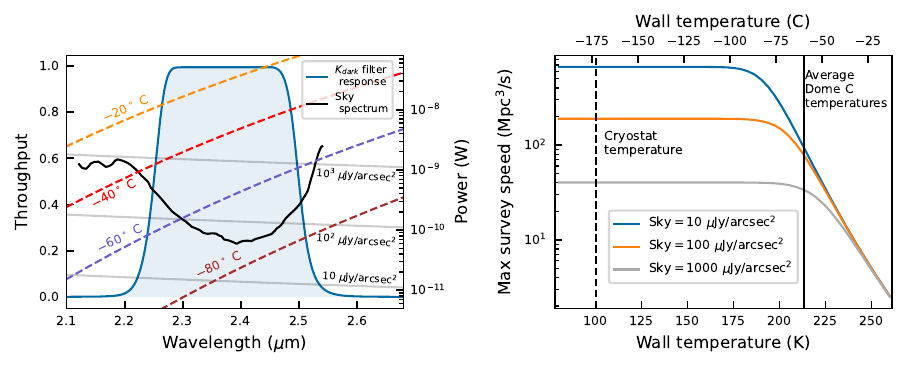}
  \caption{ {\it Left}: Estimated incident thermal background at the central detector pixel. Blackbody thermal self-emission from the cryostat walls is estimated as prescribed in the text from $-20^\circ$ to $-80${\degC}, marked by colored dashed lines. Fiducial constant sky brightness curves are over-plotted in gray, alongside the measured South Pole sky background from \citet{Ashley1996} and \citet{Nguyen1996} and Cryoscope's $K_{dark}$ filter response curve. {\it Right}: Maximum attainable volumetric survey speed for the Pathfinder as a function of the temperature of the cryostat walls, assuming the fiducial values for $K_{dark}$ sky brightness in the left panel. 
  }\label{fig:vspeed_temp}
\end{figure}

However, assuming blackbody emission from the sky is a crude approximation. In the sensitivity estimates in the main text, we find the power delivered to the detector from the thermal sky by integrating over the South Pole sky spectrum measured by \citet{Ashley1996} and \citet{Nguyen1996}, a conservative upper estimate for Dome C. Integrating over the sky spectrum, we directly compute the signal from the sky and compare to the estimated thermal emission from the telescope walls. The thermal background at the central detector pixel is shown in the left panel of Figure~\ref{fig:vspeed_temp} for a range of cryostat temperatures. The contribution from the South Pole thermal sky alongside fiducial constant sky brightness curves are overplotted. Only below $\sim -80${\degC} does thermal emission from the cryostat walls become sub-dominant to the South Pole sky brightness within the $K_{dark}$ bandpass. We show in the right panel of Figure~\ref{fig:vspeed_temp}, how the Pathfinder's maximum attainable survey speed depends on its interior wall temperature under skies with brightnesses of 10, 100, and 1000\,\,$\mu$Jy/arcsec$^2$. As expected, bright skies fundamentally limit sensitivity, and a cryogenic optical path provides minimal gains towards increasing survey speed. However, as skies become darker, a cryogenic optical path becomes relevant. Limiting depths under the darkest skies can be increased by two or more magnitudes and survey speed enhanced five to tenfold compared to those under nominal ambient conditions by maintaining the optics core at cryogenic temperatures. In the $K_{dark}$ window at Dome C, thermal self-emission can no longer be considered sub-dominant to the thermal sky. Cryocooling ensures that regardless of variations in Dome C sky brightness or ambient temperature fluctuations at ground level, survey speed is always maximized. The sky background, instead of the telescope's own thermal emission, is the fundamental dominant thermal background.

\section{Conversion between AB and Vega magnitudes in $K_{dark}$}\label{appendix:vega}
We derive the isophotal flux density and AB magnitude for Vega in the $K_{dark}$ filter for the Pathfinder and full-scale telescope. 
From Eq.\,9 in \citet{TokunagaVacca2005}, 
\begin{equation}
    {\rm AB} = -2.5\log(F_\nu)+8.926
\end{equation}
with $F_\nu$ expressed in units of Jy.

The isophotal flux density is defined in Eq.\,6 by  \citet{TokunagaVacca2005}, integrating over the frequency interval of the $K_{dark}$ filter.
\begin{equation}\label{eq:fnu_iso}
    F_\nu (\nu_{\rm iso}) = \frac{\int F_\nu(\nu) S(\nu) /\nu~ d\nu }{\int S(\nu)/\nu~ d\nu}
\end{equation}
where $\nu_{\rm iso}$ is the isophotal frequency and $S(\nu)$ is the total system response: $S(\nu) = T(\nu) Q(\nu) R(\nu) A_{\rm tel}$, with $T$ being the atmospheric transmission, $Q$ the product of the total optical throughput and quantum efficiency, $R$ the filter response, and $A_{\rm tel}$ the telescope collecting area. As an estimate for the conversion from $K_{dark}$ AB magnitudes to Vega magnitudes, we presume the sky transmission profile shown in Figure~\ref{fig:sky} in the main text. The optical throughput of the telescope, incorporating all losses, is 0.83. The QEs for the H2RG and SATIN detector were incorporated for the Pathfinder and full-scale, respectively, and the collecting areas for each telescope fall out of Eq.~\ref{eq:fnu_iso}. We use the Vega spectrum in $\tt{alpha\_lyr\_stis\_010.fits}$ distributed by the {\tt synphot} package \citep{synphot2018}. 

The conversion between AB and Vega in $K_{dark}$ is estimated to be: 
\begin{equation}
    {\rm AB} - {\rm Vega} = 2.04 \,\,\,\, {\rm  (Pathfinder)}
\end{equation}
\begin{equation}
    {\rm AB} - {\rm Vega} = 2.06 \,\,\,\, {\rm (Cryoscope)}
\end{equation}

with the difference driven by the different quantum efficiencies of the detectors between Pathfinder and full-scale. Other associated values are shown in Table~\ref{tab:vega}.

\begin{deluxetable}{cccccc}
\tablecolumns{6}
\tablecaption{$K_{dark}$ isophotal wavelengths, frequencies, flux densities and AB magnitudes for Vega}
\label{tab:vega}
\tablehead{
\colhead{Telescope} & 
\colhead{$\lambda_{iso}$} &
\colhead{$\nu_{iso}$} &
\colhead{$F_\lambda$} &
\colhead{$F_\nu$} &
\colhead{AB}\\
 & ($\,\mu$m) & ($\times 10^{14}$ Hz) & ($\times 10^{-10}$ W m$^{-2}$ $\,\mu$m$^{-1}$) &  ({\rm Jy}) & (mag)
}
\startdata
Cryoscope Pathfinder & 2.34 & 1.279 & 3.09 & 567.5 & 2.04\\
Cryoscope & 2.37 & 1.267 & 2.98 & 557.7 & 2.06\\
\enddata
\tablecomments{$\nu_{iso}$ and $F_\nu$ are calculated as described in the text. $\lambda_{iso}$ and $F_\lambda$ are calculated from Eq.\,5 in \citet{TokunagaVacca2005}. }
\end{deluxetable}

\bibliography{cryoscope}
\end{document}